\documentclass{aa}  
\usepackage{natbib}
\usepackage{graphicx}
\usepackage{txfonts}
\usepackage[normalem]{ulem}
\usepackage{xcolor}
\usepackage{placeins} 

\usepackage{caption}
\usepackage{placeins} 
\usepackage{wrapfig}

\definecolor{links}{rgb}{0.11, 0.67, 0.84}
\usepackage[colorlinks=true,allcolors=links]{hyperref}

\definecolor{nicegreen}{rgb}{0.09, 0.45, 0.27}

\newcommand{\rhodtfe}{$\rho_\mathrm{DTFE}$}

\begin{document} 

  \title{Evolution of cosmic filaments in the MTNG simulation}
  
  \titlerunning{Cosmic filaments in MTNG}
  
  \author{Daniela Gal\'arraga-Espinosa\inst{1} 
  \and Corentin Cadiou\inst{2}
  \and C\'eline Gouin\inst{3}
  \and Simon D. M. White\inst{1} 
  \and Volker Springel\inst{1}
  \and Rüdiger Pakmor\inst{1}
  \and Boryana Hadzhiyska\inst{4, 5}
  \and Sownak Bose\inst{6}
  \and Fulvio Ferlito\inst{1}
  \and Lars Hernquist\inst{7}
  \and Rahul Kannan\inst{8} 
  \and Monica Barrera\inst{1} 
  \and Ana Maria Delgado\inst{7}
  \and C\'esar Hern\'andez-Aguayo\inst{1}
    }

    \institute{Max-Planck Institute for Astrophysics, Karl-Schwarzschild-Str.~1, D-85741 Garching, Germany\label{inst1}
    \and
    Lund Observatory, Division of Astrophysics, Department of Physics, Lund University, Box 43, SE-221 00 Lund, Sweden\label{inst2}
    \and
    Université Paris-Saclay, CNRS, Institut d’Astrophysique Spatiale, 91405, Orsay, France\label{inst3}
    \and
    Physics Division, Lawrence Berkeley National Laboratory, Berkeley, CA 94720, USA\label{inst4}
    \and 
    Miller Institute for Basic Research in Science, University of California, Berkeley, CA, 94720, USA\label{inst5}
    \and
    Institute for Computational Cosmology, Department of Physics, Durham University, South Road, Durham DH1 3LE, UK\label{inst6}
    \and
    Harvard-Smithsonian Center for Astrophysics, 60 Garden Street, Cambridge, MA 02138, USA\label{inst7}
    \and
    Department of Physics and Astronomy, York University, 4700 Keele Street, Toronto, ON M3J 1P3, Canada\label{inst8}
    \\
\email{danigaes@mpa-garching.mpg.de}}

\date{Received XXX; accepted YYY}

  \abstract{
   {
    We present a study of the evolution of cosmic filaments across redshift with an emphasis on some important properties: filament lengths, growth rates, and radial profiles of galaxy densities.
    Following an observation-driven approach, we built cosmic filament catalogues at $z=0,1,2,3,$ and 4 from the galaxy distributions of the large hydro-dynamical run of the MilleniumTNG project. We employed the extensively used DisPerSE cosmic web finder code, for which we provide a user-friendly guide, including the details of a physics-driven calibration procedure, with the hope of helping future users. We performed the first statistical measurements of the evolution of connectivity in a large-scale simulation, finding that the connectivity of cosmic nodes (defined as the number of filaments attached) globally decreases from early to late times. The study of cosmic filaments in proper coordinates reveals that filaments grow in length and radial extent, as expected from large-scale structures in an expanding Universe. But the most interesting results arise once the Hubble flow is factored out. We find remarkably stable comoving filament length functions and over-density profiles, showing only little evolution of the total population of filaments in the past $\sim 12.25$ Gyrs. However, by tracking the spatial evolution of individual structures, we demonstrate that filaments of different lengths actually follow different evolutionary paths. 
    While short filaments preferentially contract, long filaments expand along their longitudinal direction with growth rates that are the highest in the early, matter-dominated Universe. Filament diversity at a fixed redshift is also shown by the different ($\sim 5 \sigma$) density values between the shortest and longest filaments. Our results hint that cosmic filaments can be used as additional probes for dark energy, but further theoretical work is still needed.
    } }


\maketitle
\newpage

\section{Introduction}

The first observations of galaxy distributions around the Perseus cluster \citep{Joeveer1978_Perseus}, around the Coma/A1367 supercluster \citep{Gregory1978_coma}, and in the Center for Astrophysics (CfA) galaxy survey \citep{Lapparent1986} showed that on the largest scales, matter in the Universe is organised in clusters, filaments, walls, and voids. These structures form the cosmic web \citep{White1987,Bond1996}, a multi-scale network of dark matter (DM) and gas, which emerged under the action of gravity from the anisotropic collapse of initial fluctuations of the density field \citep{Zeldovich1970}. Following its discovery, the cosmic web as traced by galaxies has been observed in a variety of galaxy surveys, such as the Sloan Digital Sky Survey \citep[SDSS, ][]{York2000}, the two degree Field Galaxy Redshift Survey \citep[2dFGRS,][]{Colless2003_2dFsurvey}, Cosmic Evolution and COSMOS2015 \citep{Scoville2007_COSMOSsurvey, Laigle2016cosmos}, the 6dF Galaxy Survey \citep[6dFGS,][]{Jones2009_6dFGSurvey}, the Galaxy and Mass Assembly survey \citep[GAMA,][]{Driver2011_GAMAsurvey}, the VIMOS VLT deep survey \citep[VVDS,][]{Lefevre2005}, VIPERS \citep{Guzzo2014}, WISExSuperCOSMOS \citep{Bilicki2016}, and the SAMI survey \citep{Bryant2015_SAMIsurvey}, each with a different resolution and statistical power.

While the nodes of the web, traced by the most massive clusters of galaxies, have been intensively characterised \citep[e.g.][among many others]{Navarro1997_NFW, Nagai2007, Arnaud2010, Baxter2017_splashback, Bartalucci2017, PintosCastro2019, Ghirardini2019}, cosmic filaments are still only poorly understood. This is principally due to their low matter densities which (combined with their complex morphology) result in a fainter signal with respect to that of the cosmic nodes. For example, the milestone analysis of \cite{Cautun2014} has shown that the probability distribution function (PDF) of the DM over-density field in filaments is, on average, two orders of magnitude lower than that in nodes, although the exact number is definition dependent \citep[e.g.][]{Busch2020_TLT}. Observing cosmic filaments in galaxy surveys is thus extremely challenging.

In addition to these observational difficulties, the lack of an unequivocal definition of cosmic filaments complicates the study of these structures even further. As illustrated by the detailed review of \cite{Libeskind2018}, a variety of methods have been developed to identify filaments in observational and simulated datasets, each adopting a different approach. For example, some algorithms define filaments based on the topology of the matter density field \citep{Sousbie2011b, Disperse_paper1}, on the phase-space caustics of DM \citep{Neyrinck2012_Origami}, or through the relative strength between eigenvalues of the Hessian matrix of the density field \citep[e.g.][]{AragonCalvo2007, AragonCalvo2010spineweb, Cautun2014, Pfeifer2022_COWSfilaments}. Filaments can also be identified based on statistical representations of stochastic processes \citep[e.g.][]{Stoica2007, Tempel2016bisous}, using graph theory \citep[e.g.][]{Pereyra2020_semita}, regularised minimum spanning trees \citep[e.g.][]{Bonnaire2020Trex}, or via machine learning classification tools \citep[e.g.][]{Buncher2020, Inoue2022, Awad2023}. This diversity of filament finders underlines the high complexity involved in the study and physical characterisation of these structures.

Importantly, studies of numerical simulations have shown that cosmic filaments today contain more than $50 \%$ of the matter in the Universe \citep{Cautun2014, GaneshaiahVeena2019_CosmicBalletII}. The understanding of matter at the largest scales is therefore inevitably linked to that of these structures. The last decade has thus witnessed an outstanding number of studies aiming to characterise the properties of cosmic filaments themselves at $z \sim 0$ \citep[e.g.][]{Colberg2005, GhellerVazza2015, GhellerVazza2016, Ho2018, Connor2018, GalarragaEspinosa2020, Rost2020, Pereyra2020_semita, Malavasi2020_sdss, Wang2021NatAs_spinfils, Zhu2021_filaments_evo_z, Vernstrom2021_LOFAR, Rost2021, GalarragaEspinosa2022, Zakharova2023}, the properties of the haloes \citep[e.g.][]{AragonCalvo2007, Hahn2007b_EVOprop_haloes_LSS,Musso2018_tidesLSS, GaneshaiahVeena2018_CosmicBalletI, GaneshaiahVeena2021_CosmicBalletIII, Jhee2022_filaments} and 
galaxies living in these structures \citep[e.g.][]{Pandey2006, Tempel2014, Chen2017_fil_gal, Kuutma2017, Malavasi2017, Bonjean2018, Laigle2018, Kraljic2018, Sarron2019, Kraljic2019, LiaoGao2019, GaneshaiahVeena2019_CosmicBalletII, Santiago-Bautista2020_filaments, Bonjean2020filaments, Welker2020_sami, Kraljic2020spin, Singh2020, Gouin2020_galaxy, Song2021, Malavasi2022_spin, Kotecha2022_fils_clusters, Kuchner2022_inventoryGalaxies}, 
and those of gas around cosmic filaments \citep[e.g.][]{KlarMucket2012, Nevalainen2015, GhellerVazza2019_surveyTandNTprops_fils, DeGraaff2019, Tanimura2019, Martizzi2019a, Tanimura2020_byopic, Tuominen2021, GalarragaEspinosa2021, Biffi2022_eROSITAfils, Tanimura2020_Rosat, Tanimura2022_eRosita, Holt2022,Gouin2022_gas,Erciyes2023_whim, Lu2023}, both in numerical simulations and observations. From these studies, we have learned about the broad diversity of filaments in the late Universe: from hot and dense bridges of matter between clusters, to longer, cooler (but still $T \gtrsim 10^{5}$ K on average) cosmic filaments living in less dense environments, for example close to voids. We have understood that galaxies are pre-processed in cosmic filaments before falling into clusters, and that halo properties are affected by these large-scale environments, in particular their spins. We have reached the consensus that cosmic filaments are the hosts of most of the `missing baryons' \citep{CenOstriker2006, Shull2012} in the late Universe, with the warm-hot intergalactic medium (WHIM) absolutely dominating the gas content of these structures in simulations, although this phase is still partially elusive in today's observations.

In this context, this work aims at setting the stage for pushing those studies to higher redshifts, with the goal of building a global picture of the evolution of cosmic filaments and of matter within them. We use the outputs of the large hydro-dynamical MilleniumTNG simulation \cite[e.g.][]{HernandezAguayo2023_MTNG, Pakmor2023_MTNG} together with the filament finder code DisPerSE \citep{Disperse_paper1, Sousbie2011b}, both introduced in Sect.~\ref{Sect:Data_and_Methods}, to construct cosmic filament catalogues at $z=0,1,2,3$ and 4. The extraction procedure of these catalogues is presented in Sect.~\ref{Sect:Catalogues}, in which we also detail our robustness checks performed for the filament catalogues. The latter organically led to a first exploration of the evolution with redshift of the number of connections of massive haloes to the filaments of the web, shown in Sect.~\ref{Sect:Connectivity}. Thanks to the computation of filament evolutionary tracks in Sect.~\ref{Sect:Lengths}, we analyse and interpret crucial properties such as the evolution of filament length functions and growth rates. In Sect.~\ref{Sect:Profiles}, we present the evolution of galaxy density profiles around cosmic filaments.
Finally, we summarise and further discuss our results in Sect.~\ref{Sect:Conclusions}.

The cosmic web being a multi-scale structure, it is important to point out that this work focuses on the filaments at the largest scales of the Universe (i.e. the cosmic filaments). These connect galaxy clusters and host galaxies and small groups, as opposed to the smaller-scale  ($\sim$ kiloparsec) filaments feeding individual galaxies with fuel for star-formation \citep[the `cold strems' e.g.][]{Ramsoy2021, GalarragaEspinosa2023, Lu2023}. Throughout this paper, we adopt the values of the cosmological parameters given by \cite{Planck2015Cosmo}, that are,  $\Omega_{\Lambda,0} = 0.6911$, $\Omega_{m,0}=0.3089$, $\Omega_{b,0}=0.0486$, $\sigma_{8}=0.8159$, $n_s=0.9667,$ and $h=0.6774$. All the error bars correspond to the errors on the mean values, derived from bootstrap resampling.

\section{\label{Sect:Data_and_Methods}Data and methods}

\subsection{The MTNG simulations}

We analyse the outputs of the full physics run of the MillenniumTNG (MTNG) project\footnote{\url{https://www.mtng-project.org/}}. 
This hydrodynamical simulation combines the successful galaxy formation model of the IllustrisTNG suite \citep{Pillepich2018TNGmodel, Nelson2019_TNGdata_release} with the large volume $(500 \, \mathrm{cMpc}/h)^3$ or $(738.1 \, \mathrm{cMpc})^3$ of the iconic Millennium simulation \citep{Springel2005_millenium}. Compared to the widely used TNG300 simulation, the MTNG hydrodynamical run is $\sim 15$ times larger in volume and has a mass resolution that is only reduced by a factor of 2.8. Indeed, with $4320^3$ DM particles and the same number of gas cells, the DM resolution of this simulation is $m_\mathrm{DM} = 1.12 \times 10^8 \, \mathrm{M}_\odot /h$, and the target baryonic mass is $m_\mathrm{target} = 2.0 \times 10^7 \, \mathrm{M}_\odot/h$. 
The MTNG simulation is thus ideal to analyse the interplay between the large-scales of the cosmic web and the non-linear scales of galaxy formation and evolution. In particular, it is one of the best simulations up to date to perform statistical studies of cosmic filaments, another one being the Flamingo project \citep{Schaye2023_Flamingo}. 
Further information on the MTNG project including details on the model and available outputs are provided in the introductory papers of \cite{HernandezAguayo2023_MTNG} and \cite{Pakmor2023_MTNG}, the latter with particular emphasis on the hydrodynamical run used in this work. 
The other introductory papers of the MTNG project cover a wide range of astrophysical and cosmological topics, such as the description of the lightcones \citep{Barrera2023_MTNG}, clustering analyses \citep{Bose2023_MTNG}, inference of cosmological parameters \citep{Contreras2023_MTNG}, galaxy intrinsic alignments \citep{Delgado2023_MTNG}, weak-lensing convergence maps \citep{Ferlito2023_MTNG}, halo occupation models \citep{Hadzhiyska2023_MTNG_hod, Hadzhiyska2023_MTNG}, and high redshift ($z>8$) galaxies \citep{Kannan2023_MTNG}.

The detection and analysis of cosmic filaments in this work makes extensive use of the catalogues of galaxies and groups found respectively at the centres of the SUBFIND-HBT subhaloes and Friend-of-Friends (FoF) haloes, identified on-the-fly following \cite{Springel2021}.
We perform our study at redshifts $z=0, 1, 2, 3$, and 4. These output times were chosen because their corresponding snapshots offer the full particle data, thus enabling future studies of, for example, gas properties around the filaments introduced in this work. We note that we made the conservative choice of stopping the analysis at $z=4$, as galaxy properties at higher redshifts are expected to be affected by the non-uniform UV background due to the epoch of reionisation. Finally, this work analyses the distributions of objects in real space and so does not account for peculiar velocity effects.

\subsection{\label{SubSubSect:Disperse101} The DisPerSE filament finder: A user's guide}

\begin{figure}
    \centering
    \includegraphics[width=0.5\textwidth]{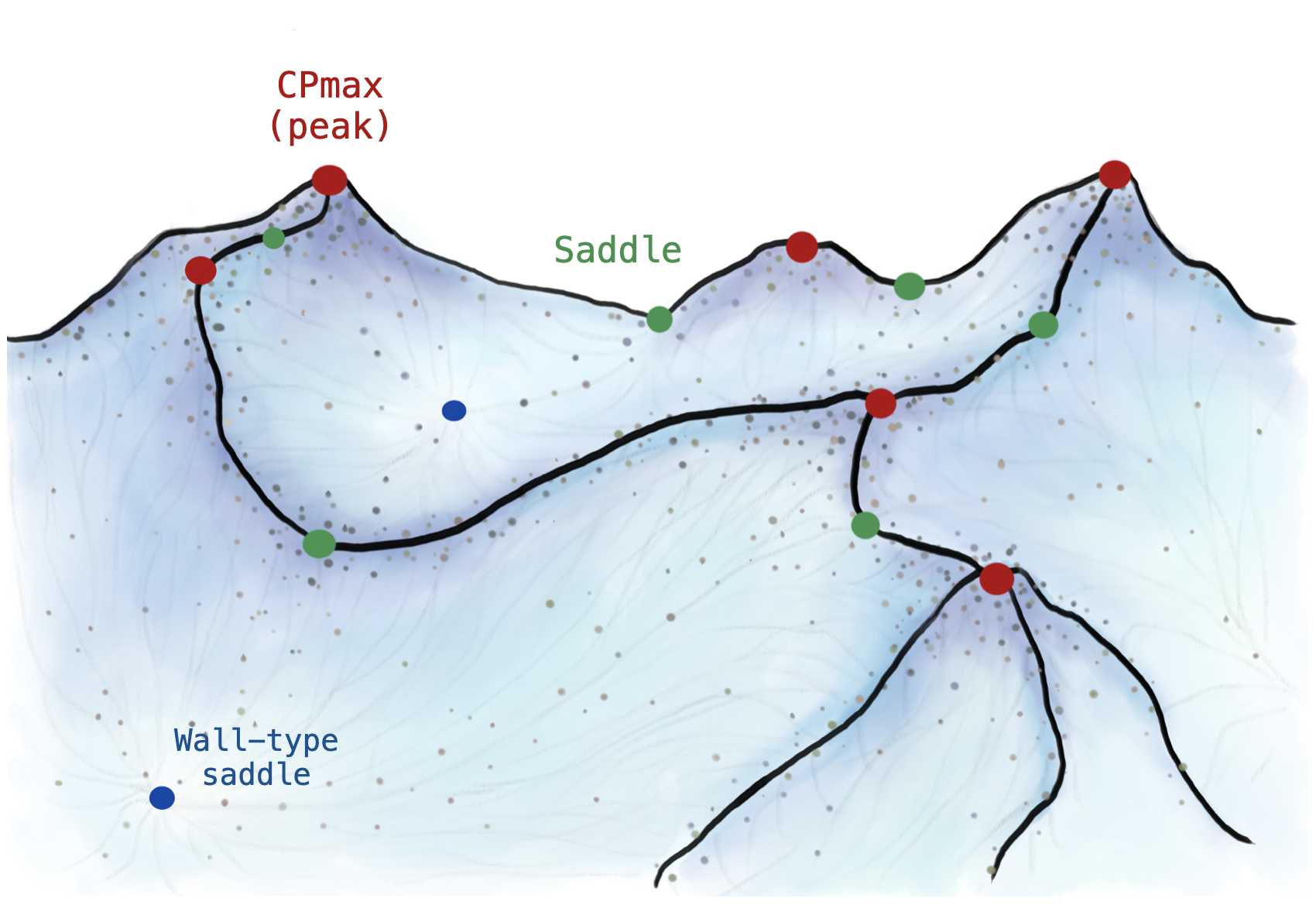}
    \caption{Sketch representing the large-scale density field and its topology using mountains as an analogy. The black lines following the ridges of the field represent the cosmic filaments. The red and green circles show, respectively, the positions of the CPmax, or peaks, and the (filament-type) saddles of the field. These are the endpoints of filaments within the DisPerSE framework. The blue circles locate the wall-type saddles of the field. Minimum density critical points are not represented.
    }
    \label{Fig:drawing1}
\end{figure}

In the following, we concisely describe the steps required for the extraction of filaments from galaxy catalogues using the DisPerSE \citep{Disperse_paper1, Sousbie2011b} cosmic web finder. With the goal of providing a complete procedure that could be helpful for future DisPerSE users, we focus explicitly on the practical details related to the application of the code. For the theoretical background at the core of DisPerSE, we refer to the presentation papers \citep{Disperse_paper1, Sousbie2011b} and website\footnote{\url{http://www2.iap.fr/users/sousbie/web/html/indexd41d.html}}. We recall that this work adopts an observation-driven approach, in which the large-scale density field is traced by the positions of galaxies (more details in Sect.~\ref{SubSect:Choiceoftracers}).

The first step for extracting DisPerSE filaments from a distribution of galaxies is to estimate the underlying (continuous) density field. This is done by the Delaunay tessellation field estimator \citep[hereafter DTFE,][]{SchaapWeygaert2000, WeygaertSchaap2009} implemented in the \texttt{delaunay\_3d} function of DisPerSE. This function provides an estimate of the density at the position of each tracer of the input dataset, which are located by definition at vertices of the Delaunay tessellation. The user can then decide to `smooth' the resulting DTFE densities (hereafter \rhodtfe) by applying the \texttt{netconv} function to the tessellation. The latter replaces the \rhodtfe~value of a given vertex by the average density of the surrounding vertices in the tessellation, thus effectively reducing the contamination by shot noise in the density values \citep[e.g.][]{Malavasi2020_sdss, Malavasi2020_coma}. The number of iterations of this averaging procedure can be decided by the user (using the \texttt{N1} parameter, see below).

The identification of the filaments is then performed by the \texttt{mse} function (the main core of the DisPerSE code). This step identifies the critical points in the (possibly smoothed) Delaunay density field, which are locations where the density gradient vanishes. Filaments are then defined as the sets of segments connecting maximum-density critical points (hereafter CPmax, or peaks) to saddles, following the ridges of the density field.
The most relevant topological features identified by DisPerSE are illustrated in the drawing of Fig.~\ref{Fig:drawing1}. In that sketch, the definition of the filaments (black lines) can be clearly understood. 
Importantly, the significance level of the detected filaments with respect to the noise can be controlled by adjusting the persistence ratio of the CPmax-saddle critical point pairs (i.e. of the filament's ending points). This parameter, hereafter \texttt{N2}, is defined as the ratio of the density of the two critical points in the pair \citep{Sousbie2011b}, and needs to be rigorously calibrated as we show in the following section.

Lastly, the \texttt{skelconv} function allows a conversion of the output of \texttt{mse} from binary to human readable format. The user can also choose to soften the angles between contiguous filament segments, exclusively for aesthetic reasons, without changing the position of the filaments' sampling points. Reading the final output of DisPerSE needs additional post-processing codes. Those used in this paper are the same as in \cite{Garaldi2023_thesan} and are publicly available\footnote{\url{https://github.com/EGaraldi/disperse_output_reader/}}. 

The different DisPerSE keywords needed for the extraction of filaments are also explicitly provided in the following. These are \texttt{--btype periodic}, \texttt{--smoothData field\_value N1}, \texttt{--forceLoops --robustness --manifolds --upSkl -nsig N2}, and \texttt{--to NDskl\_ascii -smooth 1}, respectively, for the \texttt{delaunay\_3d}, \texttt{netconv}, \texttt{mse}, and \texttt{skelconv} functions. Here \texttt{N1} and \texttt{N2} represent the only two parameters of the code, corresponding to the smoothing value of the Delaunay densities and the persistence ratio of the critical point pairs. A way of calibrating these parameters is introduced in Sect.~\ref{SubSubSect:Dispersecalibration}.
Finally, we mention that in this work we followed additional procedures in order to clean the DisPerSE outputs. We discarded filaments with zero lengths, which appear when the two ending points are located at the same position, and CPmax when associated to only one filament (together with the corresponding filament). The latter was motivated by the prior that nodes connected to a single filament are not expected at the scales of the cosmic web \citep{Codis2018}.

\section{\label{Sect:Catalogues}Extraction of cosmic filament catalogues at different redshifts}

This section presents in detail the building of the cosmic web filament catalogues using DisPerSE on galaxy distributions. We first address the selection of the tracers of the cosmic skeletons at different redshifts in Sect.~\ref{SubSect:Choiceoftracers}.
Then, in Sect.~\ref{SubSubSect:Dispersecalibration}, we explain the calibration procedure of DisPerSE parameters at $z=0$. Finally, the extraction of the higher-$z$ filament catalogues is presented in Sect.~\ref{SubSubSect:Disperse_on_high_z}.

\subsection{\label{SubSect:Choiceoftracers}Choice of tracers}

\begin{figure}
    \centering
    \includegraphics[width=0.5\textwidth]{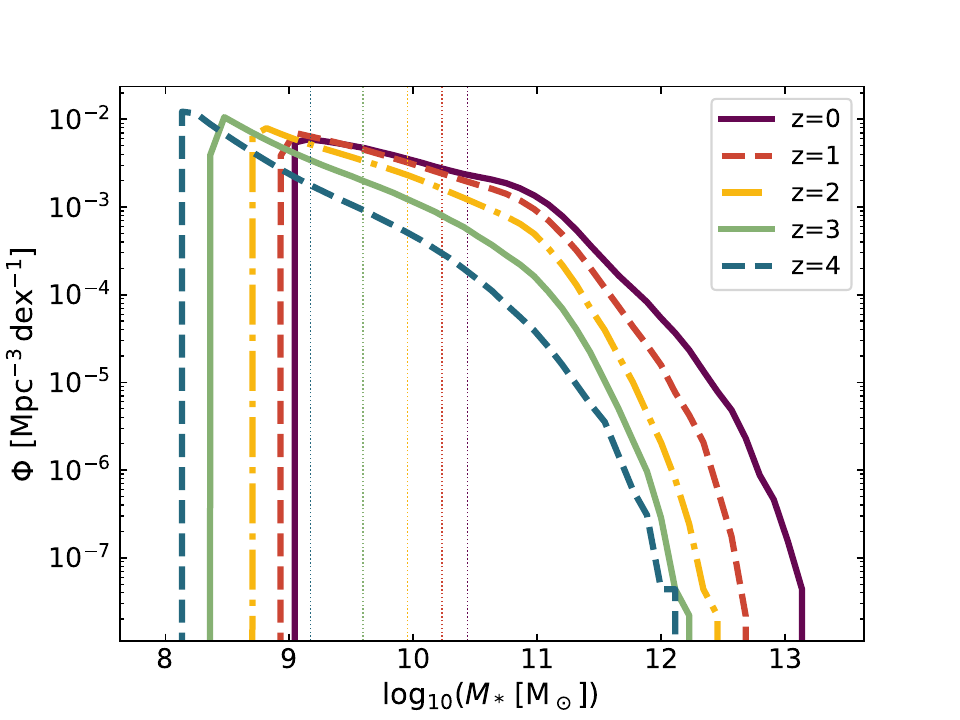}
    \caption{Stellar mass functions of the $N_\mathrm{gal}$ most massive subhaloes at redshifts $z=0$, $1$, $2$, $3$, and $4$. These objects are used to trace the cosmic skeleton. The vertical dotted lines show the mean values, i.e. $\log_{10}(M_* / \mathrm{M}_\odot) = 9.85$, $9.70$, $9.40$, $9.01$, and $8.61$, respectively from $z=0$ to $4$.}
    \label{Fig:Mstar_dist}
\end{figure}

\begin{figure*}
    \centering
    \includegraphics[width=0.58\textwidth]{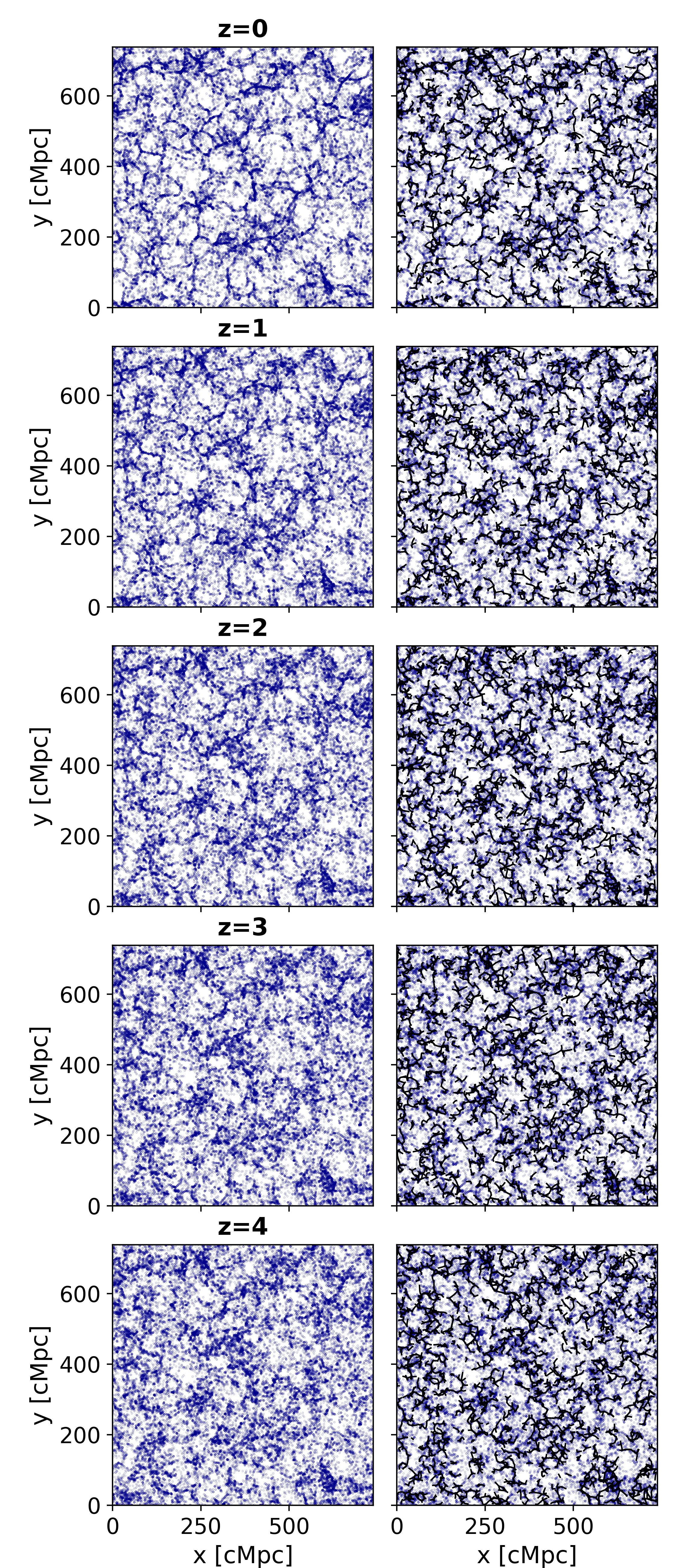}
    \caption{Slices of thickness 15 cMpc showing the galaxy distribution (left) and the associated filaments (right) at different redshifts of the MTNG simulation.}
    \label{Fig:slice_gals_fils}
\end{figure*}

At redshift $z=0$, we select from the subhalo catalogue of the MTNG simulation all the objects with stellar masses larger than $10^{9} \, \mathrm{M}_\odot$. This lower limit is chosen following the observational limits of \cite{Brinchmann2004} and \cite{Taylor2011}, and ensures a selection of subhaloes with masses large enough to trace the positions of the large-scale cosmic filaments \citep[as in e.g.][]{GalarragaEspinosa2020}. The number of objects at $z=0$ satisfying this criterion is $N_\mathrm{gal} = 3\,083\,441$, resulting in a number density of $n_\mathrm{gal} = 7.7 \times 10^{-3}$ $\mathrm{Mpc}^{-3}$ tracers in the total simulation volume. This is compatible with the expected number density of tracers in state-of-the-art galaxy surveys, such as DESI BGS at low redshift \citep{Hahn2023_DESI_bgs}. The stellar mass function of these subhaloes is presented by the dark red curve in Fig.~\ref{Fig:Mstar_dist}.

The choice of using galaxies as tracers of the density field is entirely driven by observations, where it is relatively accessible to detect the cosmic skeleton from the distribution of galaxies \citep[as done in surveys such as VIPERS, COSMOS2015, SDSS, and SAMI, respectively in][]{Malavasi2017, Laigle2018, CarronDuque2022, Malavasi2020_sdss, Welker2020_sami}.
Compared to skeletons extracted from the underlying DM distribution, this choice has a negligible impact on the positions of cosmic filaments \cite[see Appendix C of ][]{Laigle2018}, but can play a role in the number of recovered filaments, as expected from biased tracers \citep{Zakharova2023}.

At higher redshifts, the large-scale cosmic web is also traced by the most massive galaxies at a given cosmic time. We build the dataset of tracers for detecting the filaments at $z=1,2,3$, and 4, by selecting the $N_\mathrm{gal}$ most massive subhaloes (in stellar mass) in the respective SUBFIND catalogues. The same number of subhaloes as for $z=0$ is adopted precisely with the goal of working at constant (comoving) density of tracers across cosmic time. This ensures that the detected high-$z$ filaments belong to the same spatial scales as the $z=0$ cosmic filaments. 

The cosmic web being a multi-scale structure where smaller-scale objects can be found inside larger ones (e.g. galaxies connected by kiloparsecscale filaments, or tendrils inside galaxy clusters), the method adopted in this work enables the extraction and comparison between different filament catalogues without the contamination of structures at smaller-scales.
One could also think about using the first progenitors of the $\log_{10}(M_* / \mathrm{M}_\odot) \geq 9$ galaxies at $z=0$ as tracers of the higher redshift cosmic skeletons. The first reason against using them is that it would be impossible to perform the same analysis on observational data. The second is that, due to the complex pathways of mass assembly, the progenitors of massive galaxies at $z=0$ are not always the most massive objects at higher redshift. They are thus poor tracers of structure at the largest-scales, as shown in Appendix~\ref{Appendix:first_progenitors}.

The filamentary pattern is clearly present in the distribution of the most massive galaxies, visualised in the 15 cMpc thick slices of the left panels of Fig.~\ref{Fig:slice_gals_fils}. With increasing redshift from top to bottom, we can appreciate how the filamentary structure of the cosmic web is already apparent at $z=4$, and more sharply delineated at $z=0$. Finally, the stellar mass functions of the selected subhaloes at all redshifts are shown in Fig.~\ref{Fig:Mstar_dist}.

\subsection{\label{SubSubSect:Dispersecalibration}Calibration of DisPerSE at $z=0$}

\begin{figure*}
    \centering
    \includegraphics[width=1\textwidth]{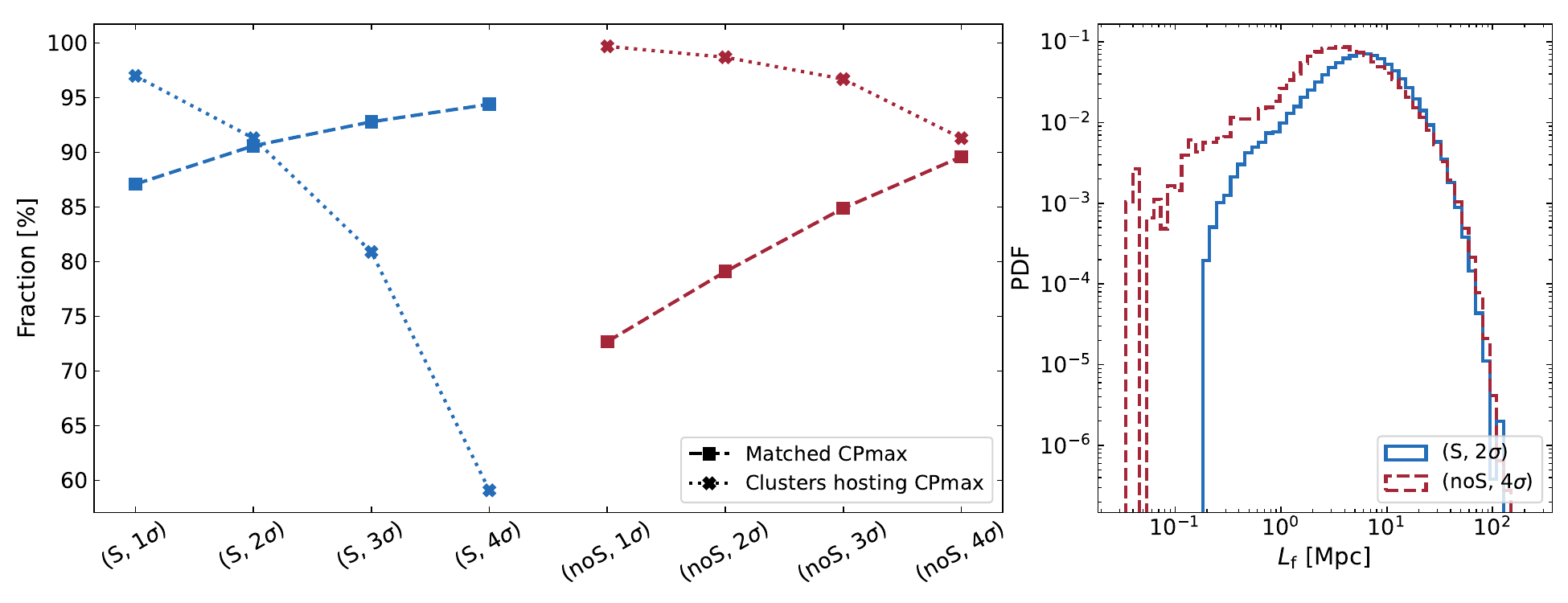}
    \caption{Calibration of DisPerSE parameters at $z=0$. \textit{Left:} Results of the matching between CPmax and massive haloes for eight different DisPerSE outputs, each with a different parameterisation (see main text). Blue and red colours show, respectively, the results for smoothed (S) and not smoothed (noS) density fields.
    \textit{Right:} Filament length distributions of the two best skeleton candidates.}
    \label{Fig:matchingCPmax_FOF__z0}
\end{figure*}

As introduced in Sect.~\ref{SubSubSect:Disperse101}, the DisPerSE filament finder possesses two main parameters that need to be carefully calibrated. Starting from the galaxy distribution at $z=0$ presented above, we perform eight different extractions of the cosmic skeleton, each with a different combination of parameters, that is, with or without smoothing of the \rhodtfe~distribution (\texttt{N1} = 0 or 1), and persistence ratios \texttt{N2} varying from $1$ to $4\sigma$. 
The best configuration of DisPerSE parameters is the one which recovers the most salient features of the large-scale distribution while minimising the effects of noise (i.e.~respectively maximising the completeness and purity of the skeleton). Nevertheless, testing the robustness of filaments is far from being a trivial task, since a `true' physical reference of the positions of filament spines can not be easily determined in simulations. To bypass this problem, one could think of applying different filament finders to the same field \citep[e.g. those of][]{Cautun2013nexus, Tempel2016bisous, Bonnaire2020Trex, Pfeifer2022_COWSfilaments} and comparing the corresponding outputs. However, this exercise would be a characterisation of the similarities and differences between the different methods \cite[as done by][]{Libeskind2018}, but would not provide conclusive results regarding the accuracy of the skeletons in detecting the actual filaments.

In this work, we assess the robustness of the skeletons by comparing the positions of the densest ending points of the filaments (the CPmax, which by construction correspond to the topological nodes) to those of the massive Friend-of-Friends (FoF) haloes of the simulation. The latter are classified into the three following mass bins: $M_{200c} \in 5 \times [10^{11}, 10^{12}]$, $M_{200c} \in 5 \times [10^{12}, 10^{13}]$, and $M_{200c} \geq 5 \times 10^{13}$ $\mathrm{M}_\odot /h$, with respectively 643\,329, 70\,392, and 6\,135 objects in the MTNG snapshot at $z=0$. Hereafter, we refer to them as galactic haloes, groups, and clusters. 
The full results of the matching between CPmax and FoF haloes can be found in Appendix~\ref{Appendix:matchingCPmax_FOF}, together with further details on each skeleton. In the following, we discuss only the most relevant information of this matching.

The first panel of Fig.~\ref{Fig:matchingCPmax_FOF__z0} shows, for each of the eight skeletons, the fraction of CPmax found within $R_{200c}$ of any of the massive FoF haloes defined above (i.e.~the matched CPmax, in dashed lines with squares), and conversely, the fraction of clusters hosting a CPmax inside their $R_{200c}$ spheres (in dotted lines with crosses). These fractions are maximised by the smoothed $2\sigma$ skeleton (S, $2\sigma$, at the intersection point in blue), and the $4\sigma$ skeleton with no smoothing (noS, $4\sigma$, in red). The other filament extractions are either too impure with too few matched CPmax caused by the lack of smoothing and low persistence ratio, or incomplete with excessively low fractions of clusters identified as CPmax because of overly high persistence values.
The right panel of Figure~\ref{Fig:matchingCPmax_FOF__z0} presents the filament length distributions of the two best candidates. We notice that the one corresponding to the `noS, $4\sigma$' skeleton presents a contribution of extremely short structures with lengths as small as 0.03 Mpc, which are likely to be originating from the noise. Therefore, we conclude that, given the $n_\mathrm{gal}$ density of tracers, the best possible extraction of large-scale cosmic filaments in MTNG is given by the smoothed-$2\sigma$ DisPerSE skeleton.

\subsection{\label{SubSubSect:Disperse_on_high_z}Higher redshift cosmic skeletons}

\begin{table*}[]
    \centering
    \begin{tabular}{l  c c c c c}
     \hline \hline
        & $z=0$ & $z=1$ & $z=1.99$ & $z=3$ & $z=4$ \\ \hline
        $N_\mathrm{CPmax}$ & 41 064 & 37 666 & 32 289 & 28 760 & 27 513\\
        $N_\mathrm{fils}$ (initial) & 169 497 & 170 352 & 153 808 & 144 539 & 143 115\\ 
        $N_\mathrm{fils}$ (final) & 155 172  ($91.5 \%$) & 154 465 ($90.7 \%$) & 134 138 ($87.2\%$) & 117 776 ($81.5\%$) & 110 539 ($77.3 \%$) \\ 
        $N_\mathrm{segs}$ (final) & 790 844 & 837 304 & 791 972 & 742 837 & 719 291 \\ \hline
        $L_\mathrm{min}$ [pMpc] & 0.20 & 0.09 & 0.10 & 0.09 & 0.08 \\
        $L_\mathrm{max}$  [pMpc] & 115.06 & 59.02 &36.41 & 34.97 & 21.00\\
        $L_\mathrm{mean}$ [pMpc] & 12.38 & 6.98 &5.32 & 4.35 & 3.64 \\
        $L_\mathrm{median}$ [pMpc] & 9.83 & 5.61 & 4.25 & 3.52 & 2.94 \\ \hline
        $L_\mathrm{min}$ [cMpc] & 0.20 & 0.18 & 0.30 & 0.37 & 0.37 \\
        $L_\mathrm{max}$  [cMpc] & 115.06 & 118.04 & 109.22 & 139.88 & 104.98 \\
        $L_\mathrm{mean}$ [cMpc] & 12.38 & 13.96 & 15.95 & 17.41 & 18.19 \\
        $L_\mathrm{median}$ [cMpc] & 9.83 & 11.21 & 12.76 & 14.07 & 14.70 \\ \hline
    \end{tabular}
    \caption{Details of the skeletons at different redshifts. Note: the length statistics are for the final sample of filaments.}
    \label{Table:filament_details}
\end{table*}

We choose to detect the high-$z$ skeletons using the same DisPerSE parameterisation as for the $z=0$ case, that is, smoothing \rhodtfe~and applying a $2\sigma$ persistence ratio between critical point pairs. This is done in order to retrieve structures that belong to the same scales and have similar density contrasts as the $z=0$ filaments. A visualisation of the resulting filaments at different redshifts can be appreciated in the right panels of Fig.~\ref{Fig:slice_gals_fils}.

\begin{figure}
    \centering
    \includegraphics[width=0.5\textwidth]{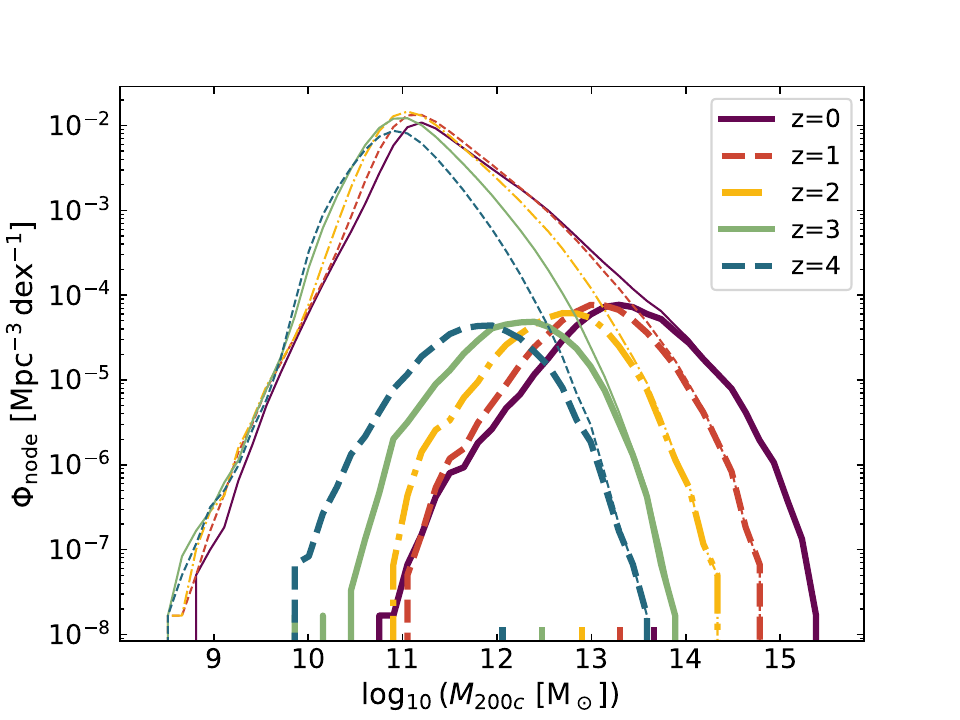}
    \caption{Mass function of the nodes connected to the filaments at different redshifts (thick curves). The vertical marks show the mean masses, which are $\log_{10}(\bar{M}_{200c}/ \mathrm{M}_\odot) = 13.66$, $13.30$, $12.90$, $12.47$, and $12.06$, respectively, from $z=0$ to $4$. For comparison, the thin curves give the mass functions of all the FoF haloes of the MTNG simulation with a stellar mass content of at least $10^8$ $\mathrm{M}_\odot$.}
    \label{Fig:mass_function_nodes}
\end{figure}

Similarly to the $z=0$ case, we check the robustness of the high-$z$ skeletons by analysing the matching between the topological nodes (the CPmax) and the FoF haloes of the corresponding MTNG snapshots. 
We find that at all redshifts the most massive haloes of the simulation are always identified as CPmax, and are thus, by construction, at the topological nodes of our skeletons. This is shown by the nice overlap towards the high mass values between the mass functions of the nodes, $\Phi_\mathrm{node}$ (in thick lines), and those of the FoF haloes (thin lines) presented in Fig~\ref{Fig:mass_function_nodes}. To be clear, we here define the mass of a given node as that of its matched FoF halo, following the procedure described in the previous section.

The number of filaments and CPmax detected at each redshift are reported in Table~\ref{Table:filament_details}. These decrease with redshift, as expected from the lower density contrasts in the early Universe compared to late times, thus resulting in fewer critical point pairs satisfying the fixed $2\sigma$ persistence ratio. 
For similar reasons, the fraction of filaments connected to massive FoF haloes also mildly decreases with redshift (from $91.5 \%$ at $z=0$ to $77.3\%$ at $z=4$). Precisely, the more homogeneous density fields at high redshifts lead to more diluted density peaks that are less likely to be identified as nodes of the cosmic skeleton.
We have checked that the disconnected filaments (i.e.~the minority of filaments whose nodes are not matched with massive FoF haloes) do not have particular features in their length distributions, density profiles, or position in the simulation box: they are just structures that connect haloes of low virial masses. Because of the cosmic scales studied in this work we choose to discard these few structures, so that the analyses of the following sections are performed on the connected filaments having more than one segment. This yields the final number of filaments reported in Table~\ref{Table:filament_details}.

\section{\label{Sect:Connectivity}Connectivity of massive haloes across redshift}

\begin{figure}
    \centering
    \includegraphics[width=0.5\textwidth]{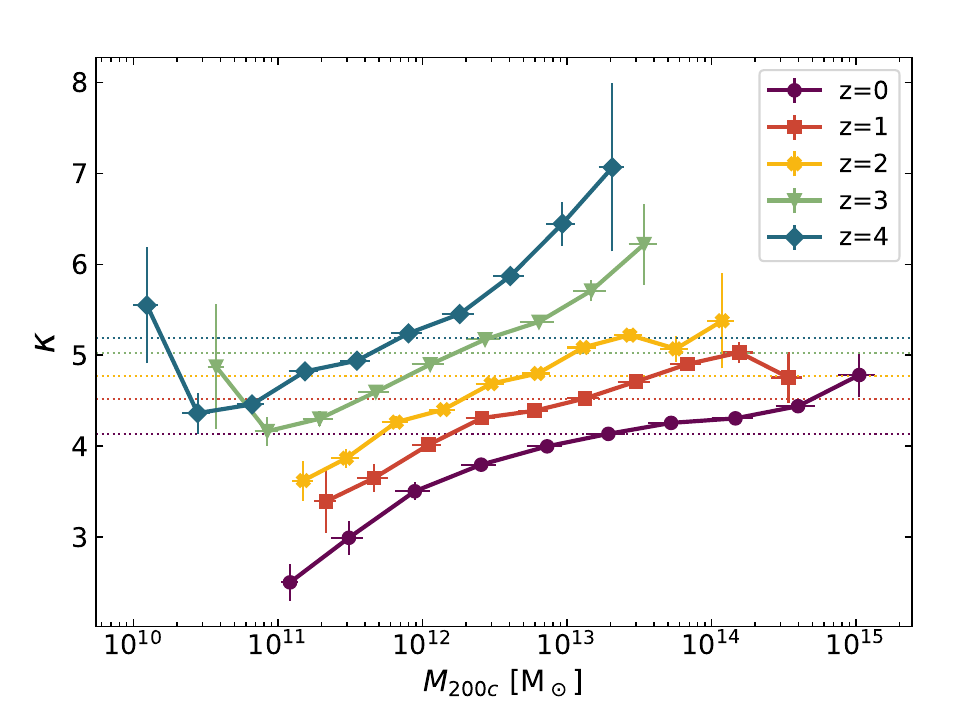}
    \caption{Evolution of the connectivity $\kappa$ as a function of halo mass. The horizontal lines show the mean connectivity values at a given redshift. From $z=0$ to 4, these are respectively 4.13, 4.52, 4.77, 5.02, and 5.19. The errorbars are derived from bootstrap resampling. }
    \label{Fig:connectivity}
\end{figure}

The cosmic filaments of the catalogues built in the previous section are all connected to massive FoF haloes (whose mass distributions were presented in Fig.~\ref{Fig:mass_function_nodes}). In this section, we perform a first analysis of the number of connections of these haloes, at the nodes of the cosmic web, with the detected cosmic filaments. Hereafter, we refer to this number as the connectivity, denoted $\kappa$.

We estimate the connectivity of a halo by measuring the number of filaments that intersect the sphere of radius $1.5 \times R_{200c}$ centred at the halo position, similar to \cite{DarraghFord2019, Gouin2021}. Our results of the mean connectivity as a function of halo mass are presented in Fig.~\ref{Fig:connectivity}. These are number-weighted averages, but we have checked that mass-weighted results are essentially the same.

At all redshifts, we find the expected trend that $\kappa$ increases with mass, in agreement with the studies of, for example, \citet{AragonCalvo2010, Codis2018, Sarron2019, Kraljic2020, LeeJaehyun2021_HR5, Gouin2021} in both simulations and observations, albeit with different normalisations due to different methods \citep[as presented in the discussion of][]{Malavasi2020_coma, Malavasi2023coma_sim}.
Interestingly, we note the presence of an `elbow' at low halo masses ($M_{200c} < 10^{11} \, \mathrm{M}_\odot$) for $z=3$ and 4, in agreement with the findings of \cite{LeeJaehyun2021_HR5} for the HR5 simulation. 

Going beyond the well know trends of halo connectivity with mass, the large statistics of MTNG and the wide redshift range explored in this work allow us to clearly show that halo connectivity consistently decreases from early to late times. Indeed, for all masses, the $z=0$ connectivity values (purple curve) are always lower than the $z=1$ ones (in red), which in turn are lower compared to the $z=2$ ones (in yellow), and so on.
These results match the theoretical predictions of \cite{Codis2018} (see their Figs.~19 and 23), and are explained by both the non-linear growth of cosmic structures (causing a decrease of $\kappa$ in the increasingly non-Gaussian fields of the late Universe), and the action of dark energy (DE). Powering the increased expansion of voids at late times, DE can also act on the merging and disconnection of cosmic filaments \citep[see e.g.][]{Cadiou2020_mergingpeaks}, thus effectively reducing the connectivity of haloes with respect to the early Universe.

This exploration of halo connectivity as a function of mass and redshift was a natural outcome of the filaments' robustness checks performed in the previous section. 
Additional studies should be performed in order to gain more insight on the evolution of connectivity across cosmic time, by, for instance, focusing on specific ranges of halo masses to better interpret the observed `elbow' at early times, or by analysing how these trends change with different DE models.
This is however beyond the scope of this paper, whose main goal is to understand the evolution of cosmic web filaments, in particular of their lengths and densities, as presented in the next sections.

\section{\label{Sect:Lengths}Filament length evolution}

\begin{figure*}
    \centering
    \includegraphics[width=0.5\textwidth]{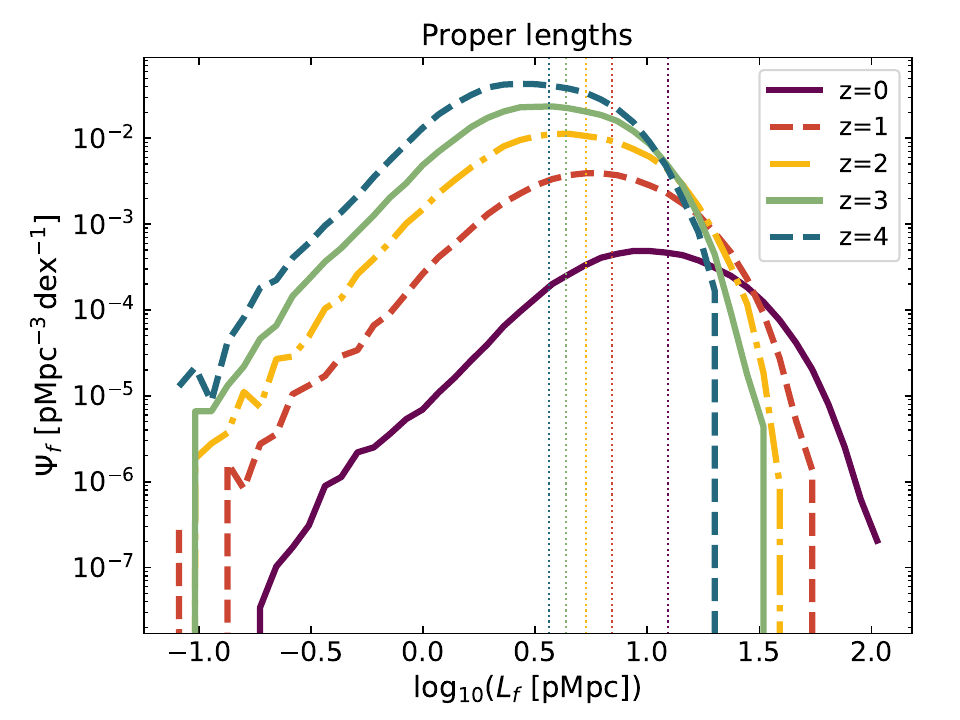}\includegraphics[width=0.5\textwidth]{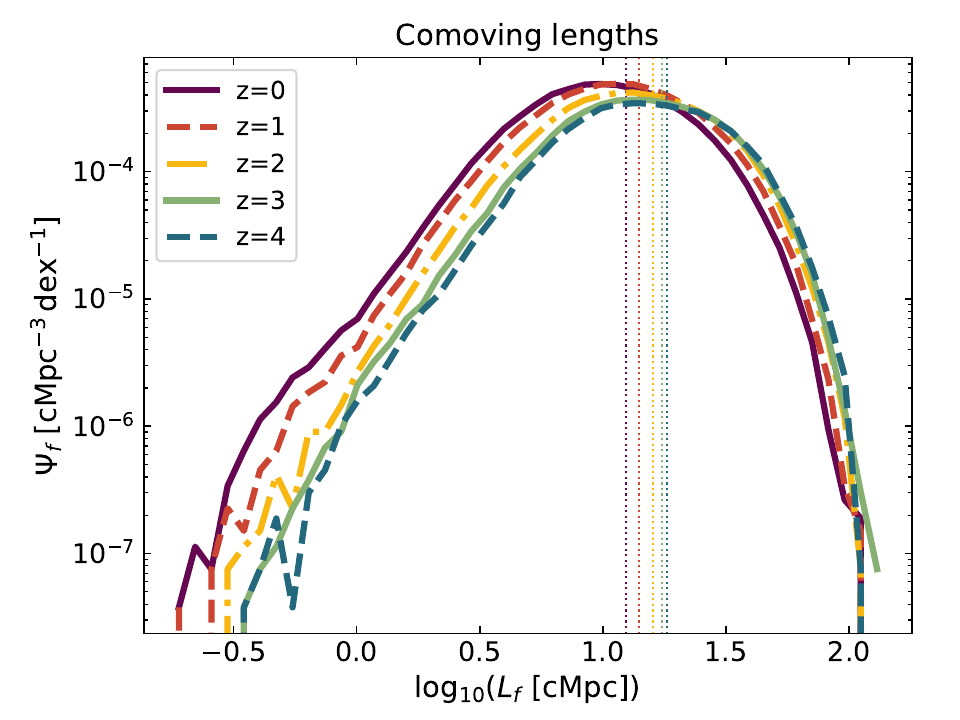}
    \caption{Filament length function $\Psi_f$ for different redshifts, in proper and comoving coordinates, in the left and right panels, respectively. The thin vertical dotted lines denote the (number-weighted) mean length, reported in Table~\ref{Table:filament_details}.}
    \label{Fig:filament_length_function}
\end{figure*}

In this section we perform a complete characterisation of filament lengths and their evolution with cosmic time. We define the length of a filament, $L_f$, as the sum of the lengths of the individual segments forming the filament, from saddle to peak.
We first introduce and analyse the filament length functions $\Psi_f$ in Sect.~\ref{Subsect:filament_length_functions}. Filament evolutionary tracks are built in Sect.~\ref{Subsect:filament_merger_trees}, and we follow the evolution of filament progenitors in Sect.~\ref{Subsect:fil_growth_rate} by estimating filament growth rates. The full length statistics of the filaments at the different redshifts are reported in Table~\ref{Table:filament_details}, in both proper and comoving coordinates.

\subsection{\label{Subsect:filament_length_functions}Filament length functions}
 
We define the filament length function, $\Psi_f$, as the number density of filaments in a bin of length rescaled by the bin width, in a similar spirit as halo mass functions. Figure \ref{Fig:filament_length_function} presents the resulting $\Psi_f$ at different redshifts, both in proper and comoving megaparsecs (hereafter pMpc and cMpc) in the left and right panels, respectively.

In proper coordinates, filament lengths increase with cosmic time. This general trend is evident both from the ranges covered by the different $\Psi_f$ and from the (number-weighted) mean values, marked by the vertical lines and reported in Table~\ref{Table:filament_details}. For example, at $z=4$ we measure a mean filament length of 3.64 pMpc while this value is 12.38 cMpc at $z=0$. This global growth in proper coordinates can be easily understood as a resut of the expanding Universe. Tied to the Hubble flow, the cosmic web expands and stretches, resulting in the general increase of cosmic filament lengths. We mention that, in proper coordinates, cosmic filaments also expand along the radial direction, increasing their radial extension with cosmic time (this is shown by the radial density profiles of Appendix~\ref{Appendix:Profiles}).

In comoving coordinates, once the Hubble flow is factored out, we can probe the specific effects of gravity and DE on the evolution of filament lengths. As revealed by the right panel of Fig.~\ref{Fig:filament_length_function}, the filament length functions are quite similar  and span overall the same ranges, indicating little evolution of the comoving lengths with cosmic time. We note, however, a slight shift of $\Psi_f$ towards high $L_f$ values at high redshift. This can be simply explained by the more homogeneous density fields at high redshifts, leading to fewer peaks (as seen in Sect.~\ref{SubSubSect:Disperse_on_high_z}), and therefore to longer filaments. 

These results characterise the lengths of all the cosmic filaments at a given cosmic time, but say little about the evolutionary pathways followed by individual structures. This question is addressed in the next sections, by building the filament histories and tracking the evolution of filament progenitors with time.

\subsection{\label{Subsect:filament_merger_trees}Filament evolutionary tracks}

\begin{figure*}
    \centering
    \includegraphics[width=0.32\textwidth]{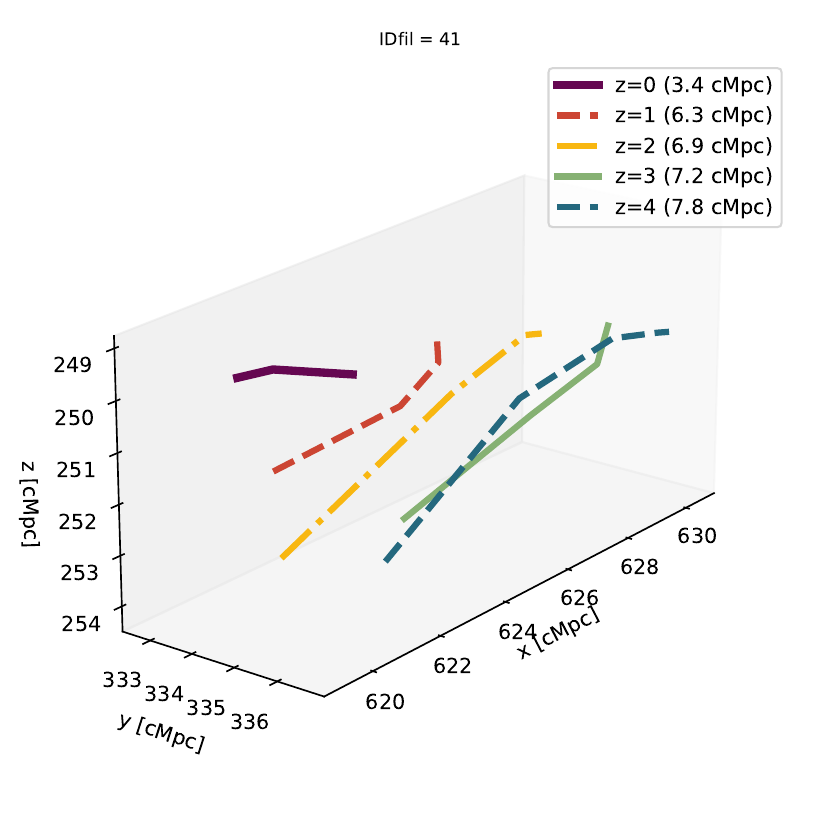}\includegraphics[width=0.32\textwidth]{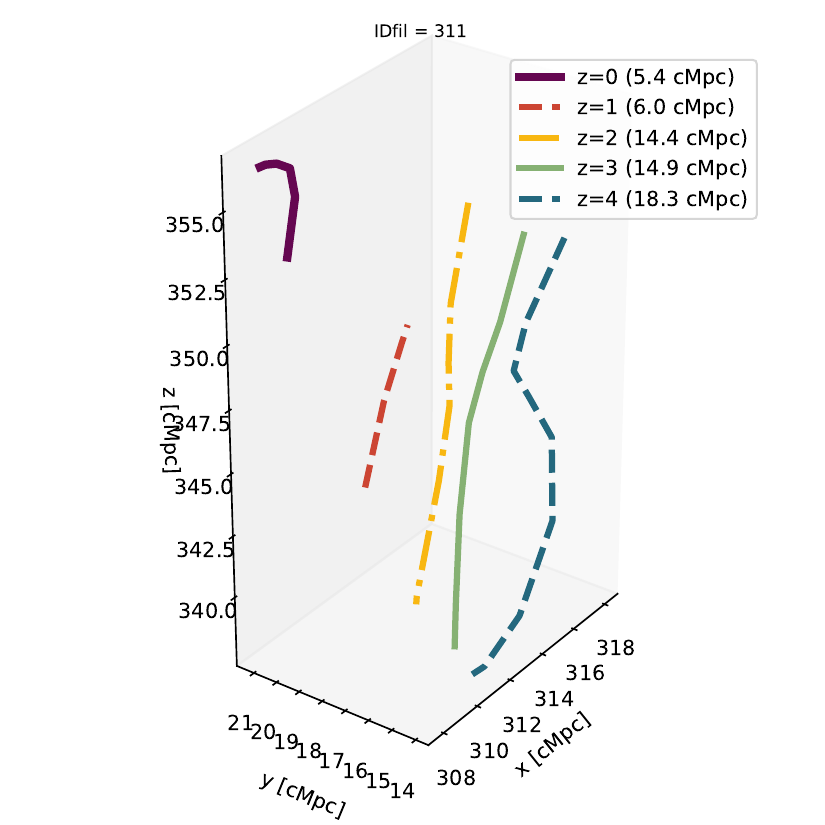}\includegraphics[width=0.32\textwidth]{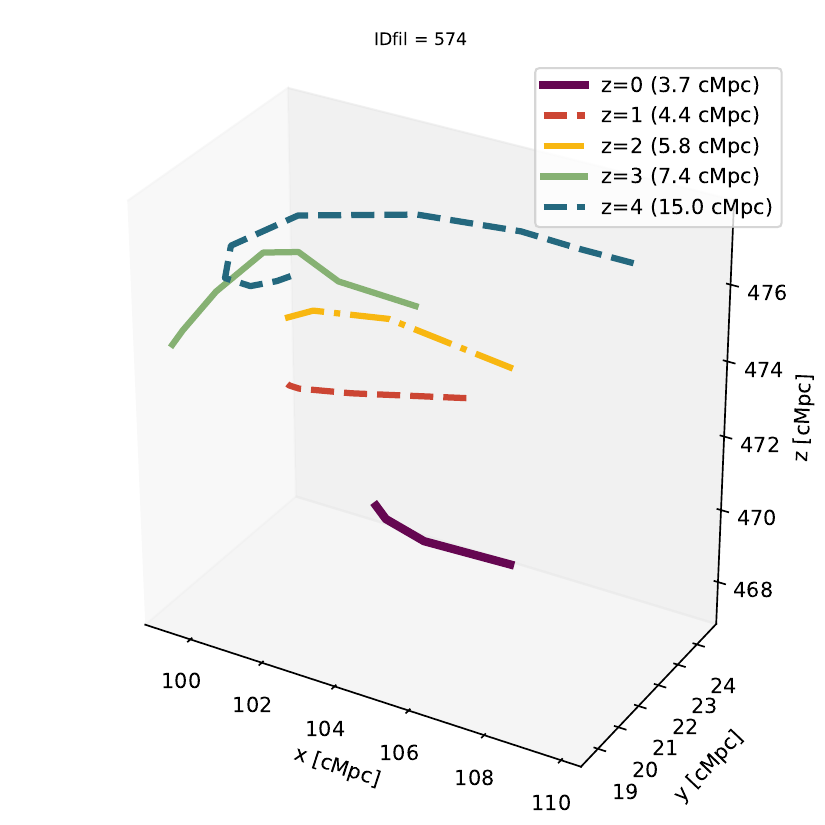}
    \includegraphics[width=0.32\textwidth]{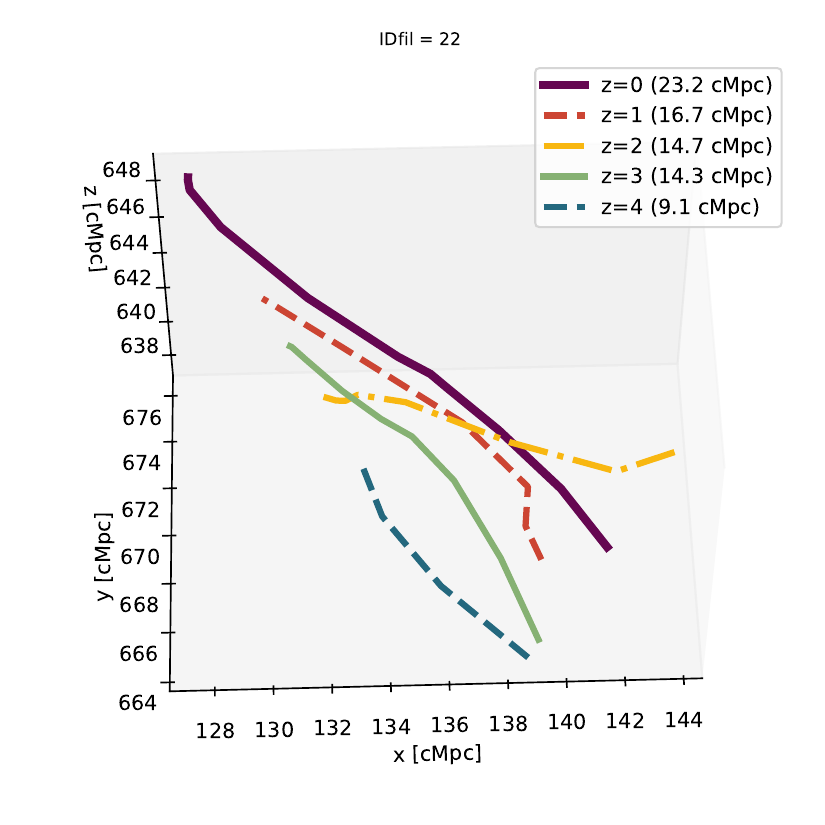}\includegraphics[width=0.32\textwidth]{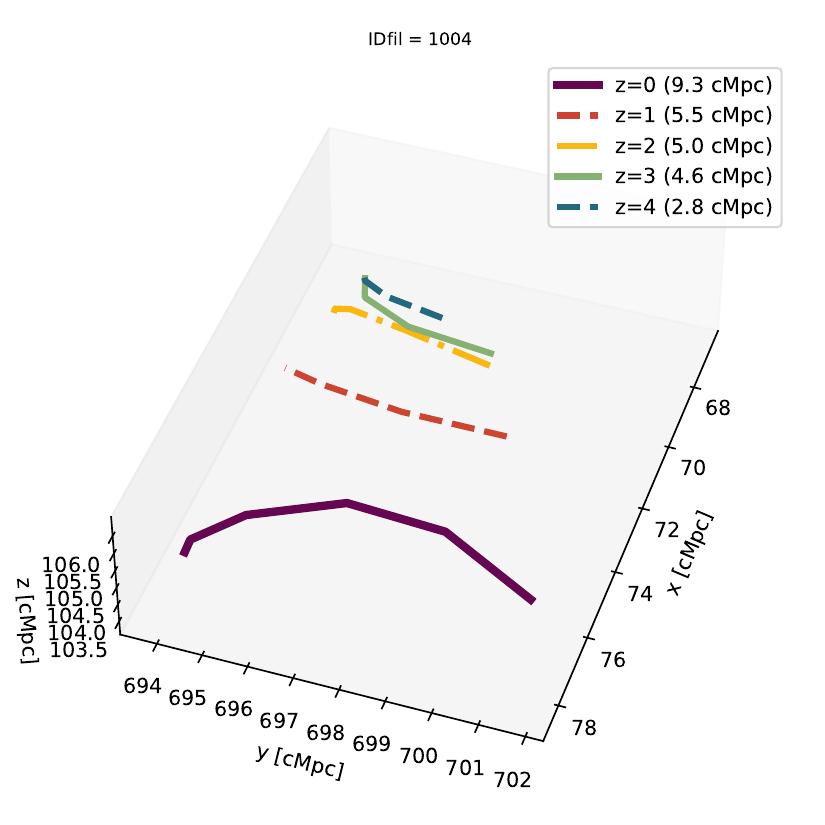}\includegraphics[width=0.32\textwidth]{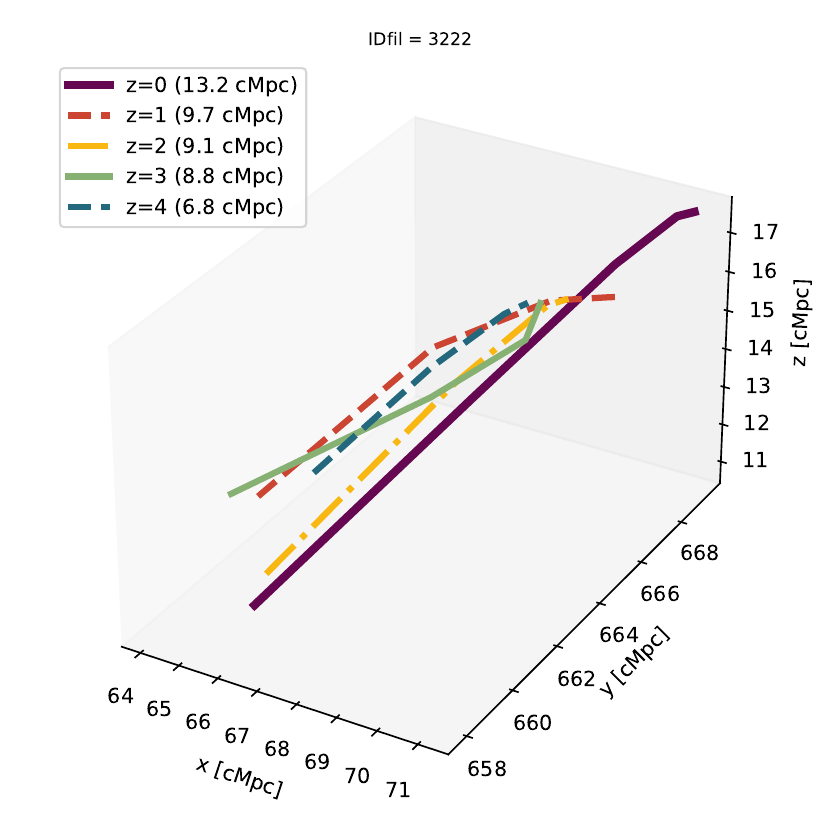}
    \caption{Example of $z=0$ filaments having complete evolutionary tracks up to $z=4$. The coloured lines represent the spines of the filaments at different redshifts. \textit{Top row:} filaments that contracted with time. \textit{Bottom row:} filaments that expanded with time. The numbers in parenthesis are the length of the filament at a given cosmic time, in comoving megaparsec.}
    \label{Fig:3d_filament_progenitors}
\end{figure*}

We gain further insights on the evolution of filament lengths by associating filaments across different snapshots. These associations aim at identifying the higher-$z$ progenitors of the $z=0$ filaments, so one can track the `same' filament back in time and explore its variations in length. In the following, we refer to these associations across cosmic time as the filament history, or evolutionary track. The code used to build them has been adapted from Cadiou et al. \textit{(in prep.)} and makes use of the scipy and numpy libraries \citep{2020SciPy-NMeth,harris2020array}.

This code works as follows:
\textit{(i)} in comoving coordinates, we compute the pairwise distances of all the peaks (i.e.~the CPmax) of the skeleton at redshift $z_1$ to the peaks of the skeleton at a higher redshift $z_2 > z_1$. We limit the computation to a distance of twice the mean nearest-neighbour separation.\footnote{This limit was set by exploring the results with different factors (from 0.5 to 4) of the mean nearest-neighbour separation. Stable results (i.e.~small variations in the final number of trees) were found with mean nearest-neighbour separations of two or higher.}
\textit{(ii)} We iterate over all pairs ($\mathrm{CPmax}_1$, $\mathrm{CPmax}_2$), ordered by increasing pairwise distance. If $\mathrm{CPmax}_1$ or $\mathrm{CPmax}_2$ have already been visited, we do nothing. Otherwise, we assign $\mathrm{CPmax}_2$ as the progenitor of $\mathrm{CPmax}_1$ and mark both as having been visited. After applying this loop to the skeletons of consecutive redshifts, we obtain lists of peaks tracked as a function of time. 
\textit{(iii)} The same two steps are applied to the saddle points $S$ of the skeletons so that we track the saddles (i.e.~the second ending points of filaments) as a function of time.
\textit{(iv)} For each filament at $z=0$ we locate its CPmax and saddle point, and we climb up their respective trunks (going to higher redshifts) as long as there exists a filament connecting them. The identified filaments are saved as the progenitors of the filament at $z=0$. 

This association produces 5061 filaments at $z=0$ with a progenitor up to $z=4$ (i.e.~with complete tracks from redshift zero to four). This is a very low number with respect to the total ($3.3 \%$), but can be interpreted by some of the processes that filaments undergo during their evolution, such as mergers, drifting, and twisting due to gravitational interactions with other structures \citep{Cadiou2020_mergingpeaks}.
Studying these dynamical processes would require a robust model of filament formation which, of our knowledge, is currently lacking in the field and goes beyond the scope of this paper.
The sample of 5061 filaments with complete histories is thus clearly biased towards structures that had little spatial evolution in the last $\sim 12.25$ Gyr (the lookback time at $z=4$). 
We have checked that the length distributions of the associated filaments show similar trends in comoving coordinates as those of the total population at their corresponding redshift. The number of filaments at $z=0$ with tracks stopping at $z=1, 2$ and 3 is, respectively, 68\,268, 31\,659, and\ 14\,052. 

For further insight, we display in Fig.~\ref{Fig:3d_filament_progenitors} some visualisations of resulting filament evolutionary tracks. For all the 3D boxes of this figure, we have plotted the $z=0$ filament (in purple) and its progenitors at higher redshifts in the same comoving box. 
The slight drift in comoving position between filament progenitors shows the mitigated effect of gravitational tides in these filaments.
In the top row, we see the example of three filaments that have followed an evolutionary path of collapse (or contraction) along the longitudinal direction. Indeed, their lengths (numbers in parenthesis) at $z=0$ are smaller than those of their progenitors at higher redshifts. The opposite behaviour of filament growth (or expansion) is illustrated in the bottom row of Fig.~\ref{Fig:3d_filament_progenitors}, where we observe the increase of progenitors' lengths from $z=4$ to 0.

\subsection{\label{Subsect:fil_growth_rate}Growth rates of cosmic filaments}

\begin{figure}
    \centering
    \includegraphics[width=0.5\textwidth]{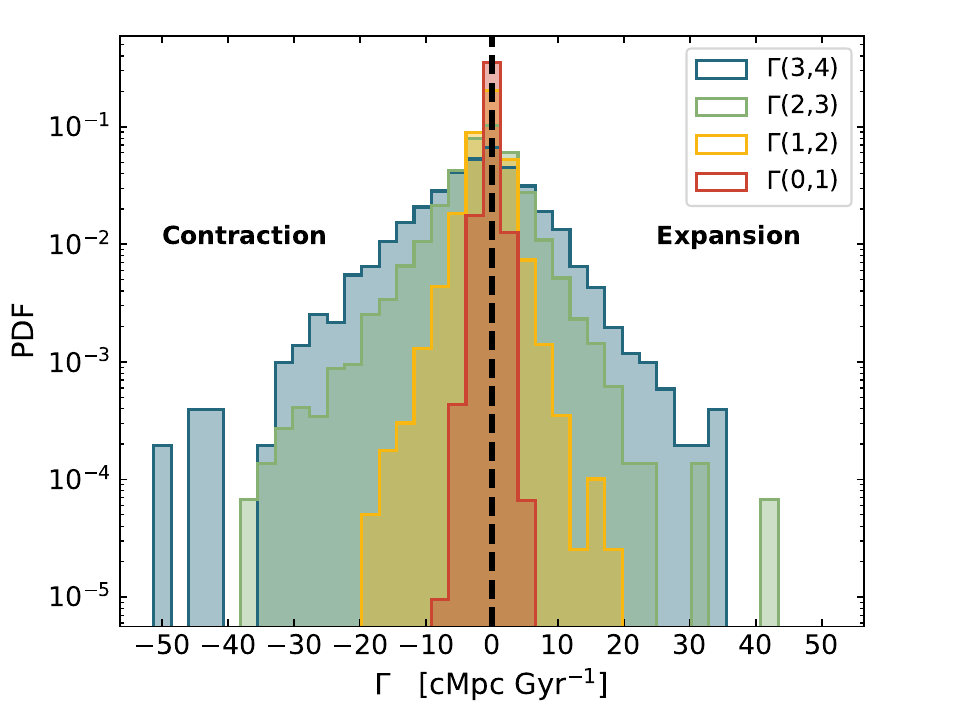}
    \caption{Growth rate distribution of cosmic filaments computed with Eq.~(\ref{Eq:gamma}). The arguments of $\Gamma(z_1, z_2)$ correspond to redshifts such that $z_2 > z_1$.}
    \label{Fig:growth_rate_distrib}
\end{figure}

\begin{figure*}
    \centering
    \includegraphics[width=0.45\textwidth]{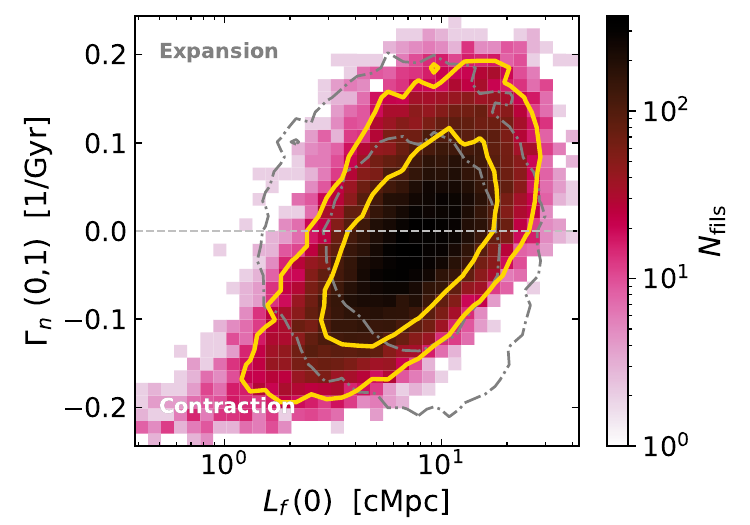}\includegraphics[width=0.45\textwidth]{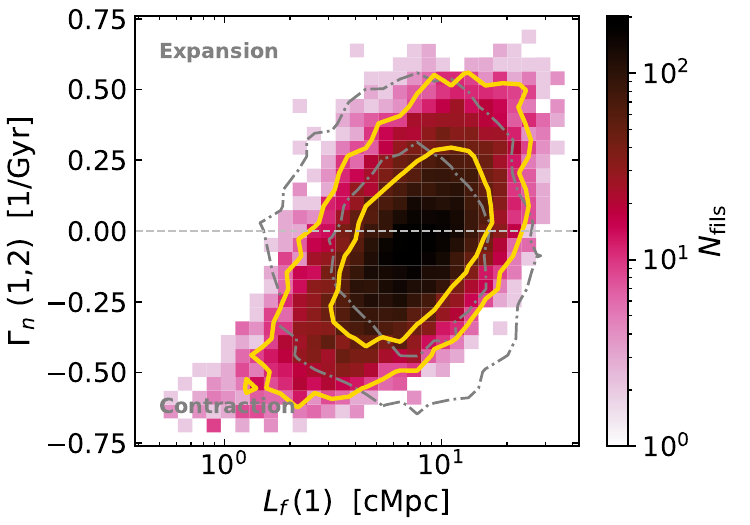}
    \includegraphics[width=0.45\textwidth]{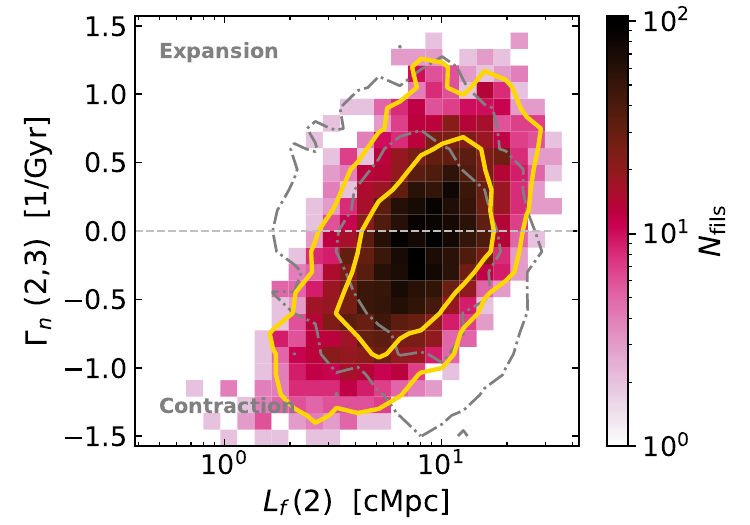}\includegraphics[width=0.45\textwidth]{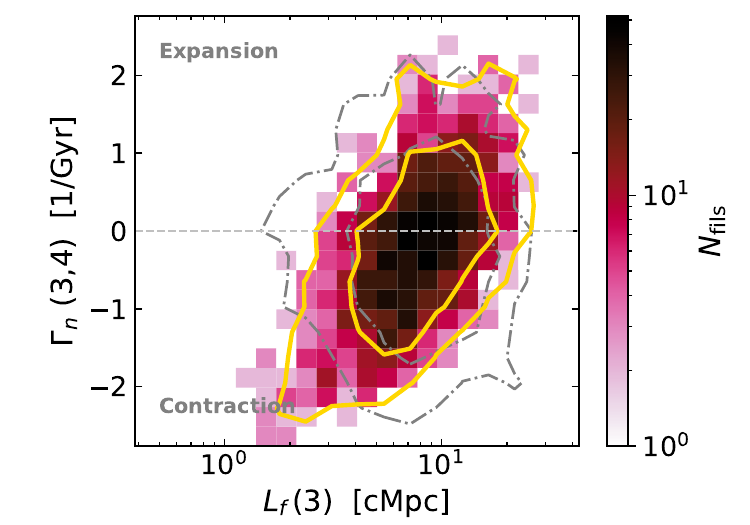}
    \caption{Normalised growth rate $\Gamma_n$ (Eq.~\ref{Eq:GammaN}) vs filament length for four consecutive redshift intervals. For each panel, the \textit{x}-axis corresponds to the filament lengths of the lowest redshift of the interval. The yellow lines indicate the $68\%$ and $95\%$ contours of the signal (pink colourbar), while the grey contours correspond to a random signal computed by associating filaments and progenitors in a random way, thus breaking the structure of the evolutionary tracks.}
    \label{Fig:growth_rate_vs_L}
\end{figure*}

We now perform a more quantitative investigation of the expansion and contraction along the longitudinal direction of filaments. We compute the filament growth rate $\Gamma$ defined as:
\begin{equation}\label{Eq:gamma}
    \Gamma(z_1, z_2) \equiv \frac{L_f(z_1) - L_f(z_2)}{| t(z_1) - t(z_2)|},
\end{equation}
where the redshifts are such that $z_2 > z_1$, the filament length $L_f$ is in units of cMpc, and $t$ is the age of the Universe in units of Gyr at a given redshift. With this definition, collapse or contraction along the longitudinal direction [i.e.~$L_f(z_1) < L_f(z_2)$] is characterised by $\Gamma <0$. Filament growth or expansion with time [$L_f(z_1) > L_f(z_2)$], is measured when $\Gamma>0$. 
Since the inputs of this quantity are lengths at consecutive redshifts, we use all the available filament progenitors presented in the previous section, even those whose history track is not complete up to $z=4$. In order to avoid any bias due to extremely wiggly filaments (which represent only a small fraction of the filaments), we take particular care in performing this analysis on relatively straight filaments. These are the structures whose ratio between length and distance between endpoints (following a straight line) is smaller than 1.2. We have checked that lower values only impact the statistics, without changing the results.

Figure~\ref{Fig:growth_rate_distrib} exhibits the distributions of growth rates in intervals of consecutive redshift [i.e.~$\Gamma(0,1)$, $\Gamma(1,2)$, etc]. Overall, these are rather symmetrically centred around zero, showing that filament contraction is as likely as expansion at all times. Interestingly, the $\Gamma$ ranges significantly narrow down at late times. For example, between redshifts three and four (blue histogram) the maximum magnitude of $\Gamma$ is $50.3$ cMpc/Gyr, while it is only $6.6$ cMpc/Gyr for $|\Gamma|$ between redshift zero and one (in red). This suggests a steadier evolution of (comoving) filament lengths in the late Universe compared to early times, when Universe was matter dominated.

\subsection{Normalised growth rates vs filament length}

Let us now explore the possible dependence of growth rate on filament length. We define the normalised growth rate, $\Gamma_n$, as the growth rate $\Gamma$ divided by the average length between the filament and its progenitor, thus giving the following full expression: 
\begin{equation}\label{Eq:GammaN}
    \Gamma_n (z_1, z_2) \equiv  
    \frac{2}{| t(z_1) - t(z_2)|} \times \frac{L_f(z_1) - L_f(z_2)}{L_f(z_1) + L_f(z_2)}.
\end{equation}
The relation between filament expansion and length needs to be explored using $\Gamma_n$ instead of $\Gamma$, since the latter is intrinsically correlated with $L_f$ (as longer filaments have higher $\Gamma$ values than short ones, by definition).

Figure~\ref{Fig:growth_rate_vs_L} presents the distribution of filaments in the $\Gamma_n - L_f$ plane. Each of the four panels of this figure shows a different redshift interval, with the \textit{x}-axis corresponding to the filament lengths of the lowest redshift. We compare the signal (pink pixels and their corresponding $68\%$ and $95\%$ yellow contours) to the results from random pairings of filaments, presented by the grey contours in the figure. While the latter are quite circular and symmetric along the $\Gamma_n=0$ axis (as expected from random associations), the former exhibit quite some deviations from this symmetry, demonstrating a positive relation between the normalised growth rate and filament length.
Indeed, the tilted and elongated yellow contours of filament progenitors shows that long filaments preferentially expand whereas the short ones contract (with redshift dependent speeds, as shown in the previous section).
In particular, we find that, at all times, the longest filaments ($L_f \gtrsim 20$ cMpc) are always expanding in comoving coordinates, and the shortest filaments of lengths $\lesssim 2.5$ cMpc are contracting, as they occupy exclusively the $\Gamma_n <0$ region.

These results demonstrate that filament expansion and contraction are related with filament length. Further characterising this dependency (by e.g. fitting models to quantify the slope of $\Gamma_n$ with $L_f$) and performing the same analysis in simulations with different DE models could help us understand how to use cosmic filaments as probes of DE. While this approach is beyond the scope of the present work, it would be extremely timely given the recent and forthcoming galaxy surveys such as DESI, Euclid, JPAS, and PFS, which will enable an unprecedented study and characterisation of the large-scale structure of the Universe.

\section{\label{Sect:Profiles}Evolution of radial density profiles}

We finally analyse the evolution of the matter distribution along the radial direction transverse to filaments. This direction is parameterised by the radial coordinate $r$, which is a measure of the perpendicular distance from a mass element to the filament spine. Taking the positions of galaxies as proxies of the distribution of matter, we compute the radial profiles of galaxy density around cosmic filaments and present them in the following.

\subsection{\label{SubSect:profiles_all}All filaments}

\begin{figure}
    \centering
    \includegraphics[width=0.5\textwidth]{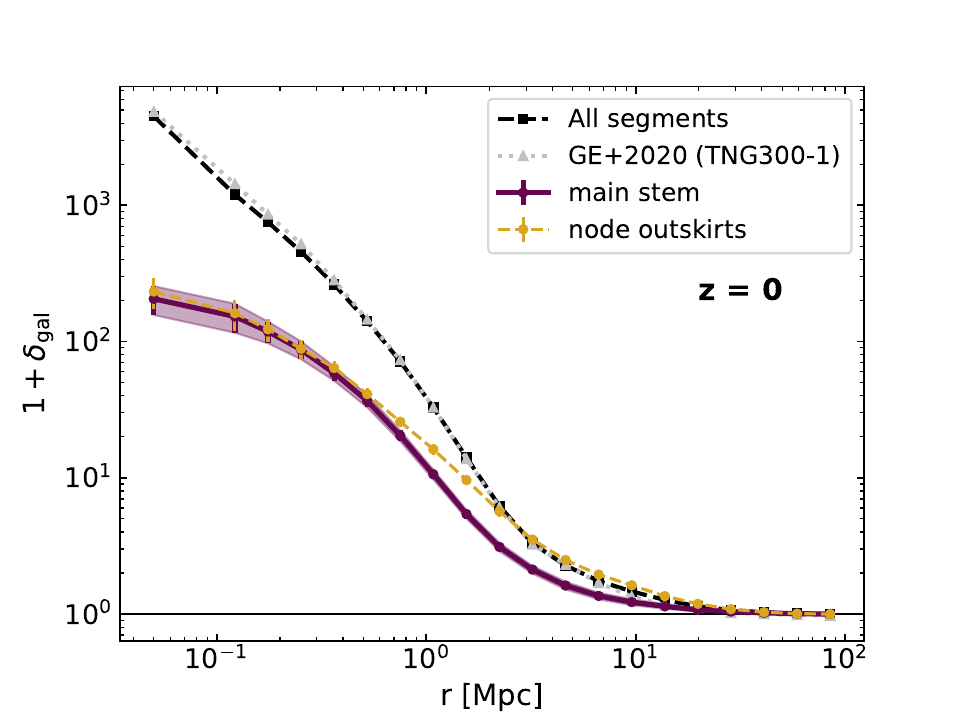}
    \caption{Radial density profile of galaxy over-densities, $1+\delta_\mathrm{gal}$, around cosmic filaments. The black dashed line corresponds to the average profile of all the filament segments, which is compared to the grey profile of \cite{GalarragaEspinosa2020}. The purple and mustard profiles correspond to the average densities in the main filament stem, and in the node outskirts, as described in the main text.}
    \label{Fig:profile_z0_comparisonTNG}
\end{figure}

\begin{figure}
    \centering
    \includegraphics[width=0.5\textwidth]{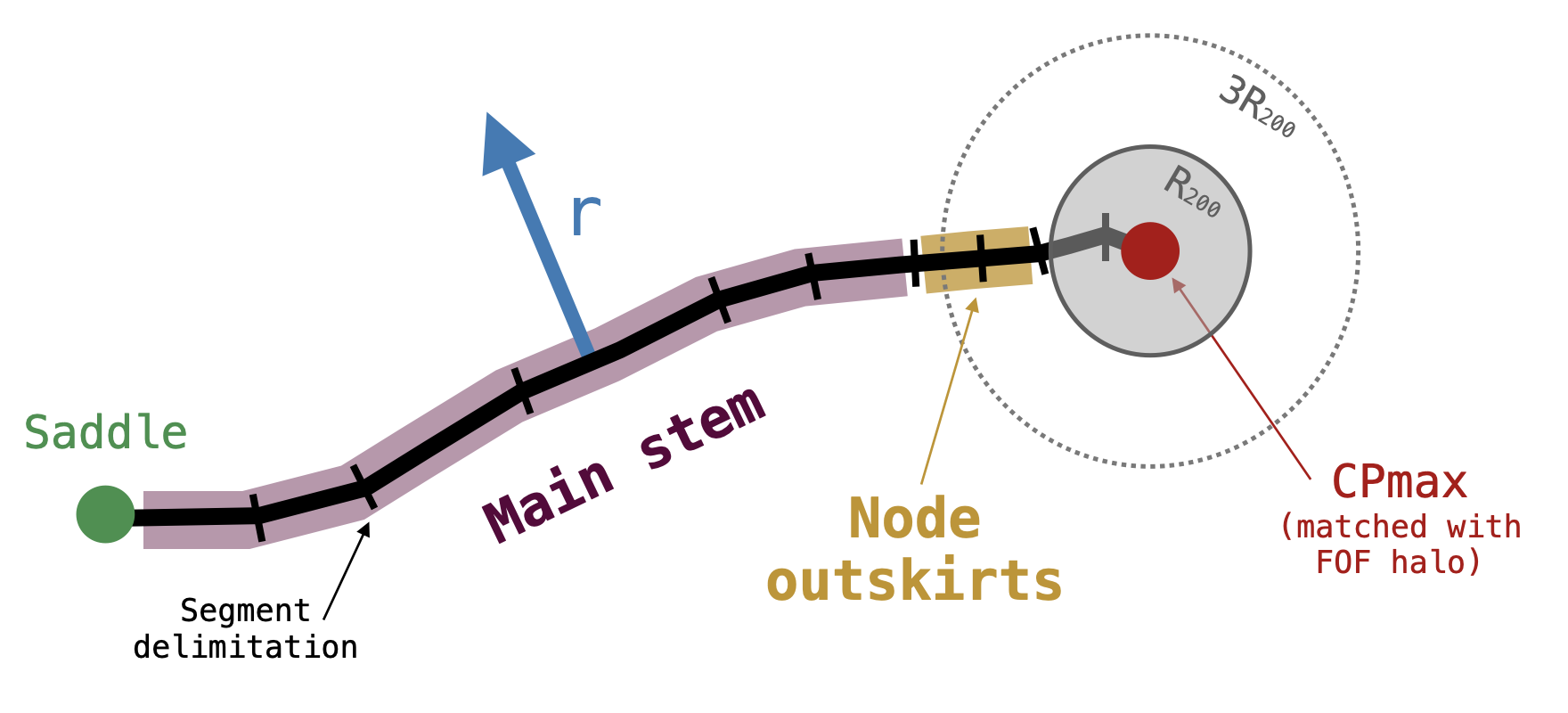}
    \caption{Sketch exemplifying a filament and its partitioning into the different regions considered for the radial density profiles. Most of the profiles of Sect.~\ref{Sect:Profiles} focus on the main stem of filaments (purple segments).}
    \label{Fig:drawing2}
\end{figure}

\begin{figure}
    \centering
    \includegraphics[width=0.5\textwidth]{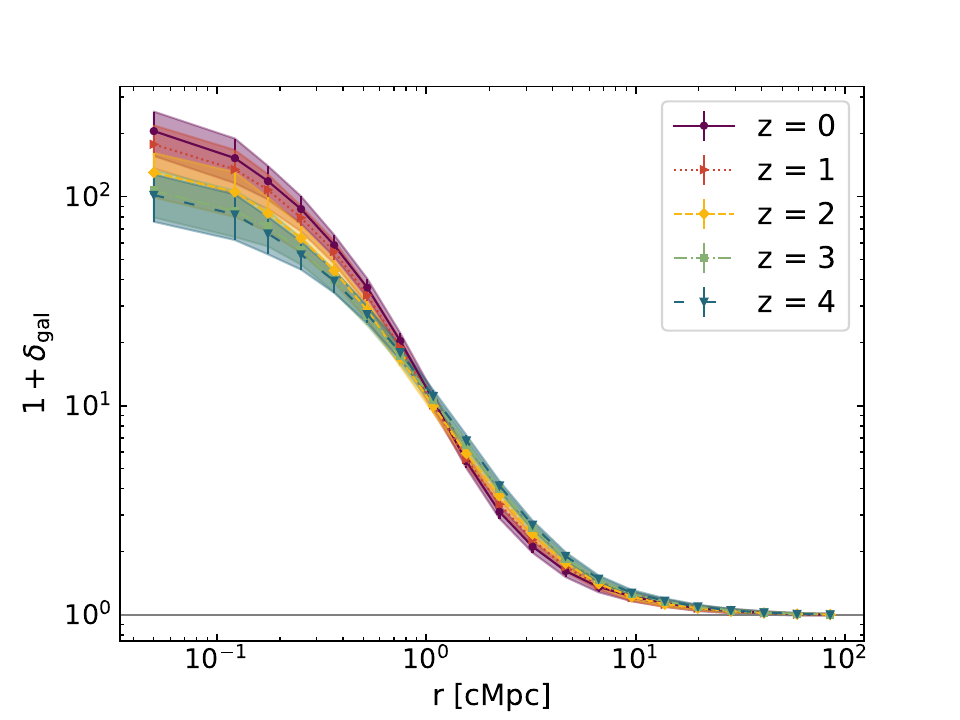}
    \caption{Radial density profile of galaxy over-densities $1+\delta_\mathrm{gal}$ around the main stem of cosmic filaments of different redshifts. The radial coordinate $r$ is in units of cMpc.}
    \label{Fig:profiles_all_z01234}
\end{figure}

Adopting the same method as in \cite{GalarragaEspinosa2020} (see their Sect.~3.3), we build the average density profiles of galaxies around the filaments presented in Fig.~\ref{Fig:profile_z0_comparisonTNG}. In a nutshell, the method consists of counting the number of galaxies in concentric hollow cylinders of increasing radii, using as axis the spine of the filament, and then dividing this number by the corresponding volume. This procedure is applied to every filament segment defined by DisPerSE (see sketch in Fig.~\ref{Fig:drawing2}), thus producing the same number of profiles as filament segments. For reference, the number of filament segments is reported in Table~\ref{Table:filament_details}.
We compute galaxy over-densities, hereafter $1+\delta_\mathrm{gal}$, by dividing the density profiles by the background density of galaxies ($n_\mathrm{gal}$, see Sect.~\ref{SubSect:Choiceoftracers}).
The average profile of all the filament segments at $z=0$ is displayed by the dashed black curve in Fig.~\ref{Fig:profile_z0_comparisonTNG}. This profile is in excellent agreement with the results of \cite{GalarragaEspinosa2020} using the TNG300-1 simulation (grey dotted curve).

Since we focus solely on the signal from cosmic filaments, we remove from the analysis the filament segments associated with the nodes of the cosmic web. In all what follows we hence discard the segments inside spheres of radius $R_{200c}$ centred at the position of the FoF haloes, at the ending points of the filaments. These are the segments in the grey zone in the sketch of Fig.~\ref{Fig:drawing2}.
The contributions from the node outskirts (defined by the $1-3 \times R_{200}$ regions) are also discarded in order to retrieve the signal of the main stem of filaments, i.e. only of the portions of filaments that are at distances larger than $3 \times R_{200}$ from the centres of the connected nodes. For the sake of clarity, Fig.~\ref{Fig:drawing2} presents a schematic illustration of the partition of a filament into these different zones. 

The average profile of segments along the main stem of filaments is plotted in purple in Fig.~\ref{Fig:profile_z0_comparisonTNG}. This curve shows that densities in the main stem are more concentrated towards filament cores than in the node outskirts (in mustard). Indeed, the latter exhibits larger radial extensions, as expected for these particular environments where filaments and clusters intersect. We note that these two profiles overlap in filament cores, at densities that are significantly lower than those obtained when the contributions from the nodes are included. In the latter case, the resulting curve (in black) is steeper and more concentrated towards small radii than both the purple and mustard curves, with a shape close to that of NFW haloes. This simply reflects how node contributions are completely dominating over filament densities when taken into account in the computation.

Now targeting the study of the main stem of filaments, we present the resulting average radial profiles at redshifts $z=0$, $1$, $2$, $3$, and $4$ in Fig.~\ref{Fig:profiles_all_z01234}.
We choose to plot these profiles as a function of the comoving distance to the spine ($r$ in units of cMpc) in order to probe trends beyond the expected growth of filaments in the radial direction, caused by the expansion of the Universe. This radial growth can be seen in proper coordinates in Appendix~\ref{Appendix:Profiles} (which also includes the versions in proper coordinates of all the figures of this section).
In Fig.~\ref{Fig:profiles_all_z01234}, it is striking to see the close similarities between the different profiles. Besides the mild increase of densities at the cores ($r < 1$ cMpc) at late times -- in qualitative agreement with \cite{Zhu2021_filaments_evo_z} and probably due to the gravitational collapse along the radial direction -- we see a large overlap at larger distances from the spines, at the outskirts of filaments ($r > 1$ cMpc) and further away. This is remarkable considering the large redshift range that is explored in this work, and hints at very little evolution of the comoving radial densities with cosmic time. This is further explored next.

\begin{figure*}
    \centering
    \includegraphics[width=1\textwidth]{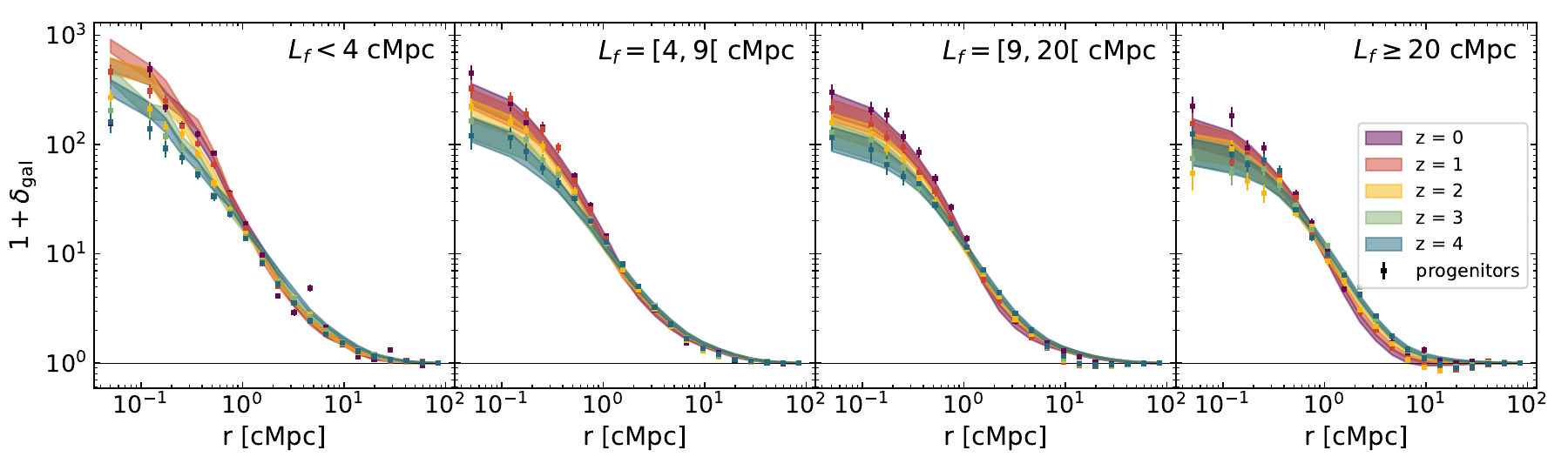}
    \caption{Same as Fig.~\ref{Fig:profiles_all_z01234} but with filaments separated in bins of comoving length, ordered from left to right by increasing lengths.}
    \label{Fig:profiles_z01234_COMbinsL}
\end{figure*}

\subsection{Profiles by bins of comoving length}

\begin{figure}
    \centering
    \includegraphics[width=0.5\textwidth]{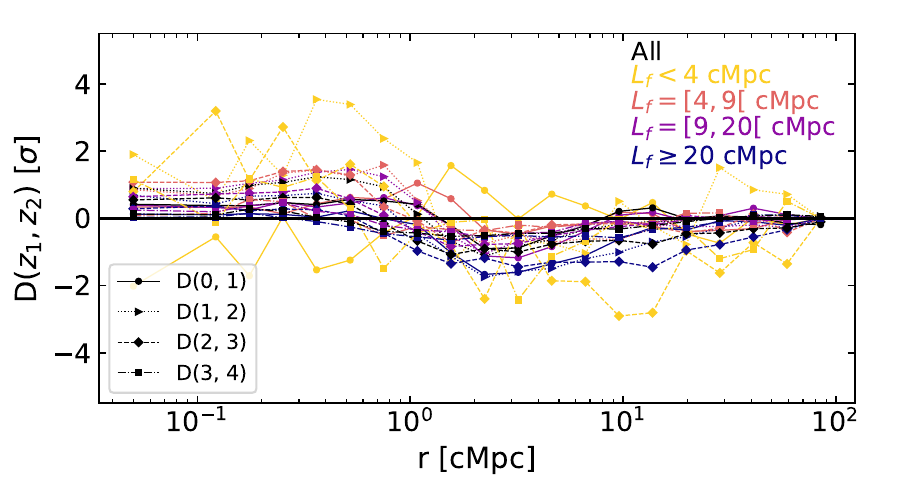}
    \caption{Differences $D$ between profiles of consecutive redshifts (Eq.~\ref{Eq:difference}), in black for all the filaments, and in other colours for filaments in a particular bin of comoving length, as labelled.}
    \label{Fig:profiles_variation_significance}
\end{figure}

Let us now split the filaments in bins of comoving length and analyse their corresponding radial density profiles. 
The selection of filaments is again done according to their comoving lengths, because we aim at probing the evolution of a given population of filaments (e.g.~the short, or the long population) across redshift. 
We compute the average profiles in the four bins of length $L_f < 4$, $[4, 9]$, $[9, 20]$, and $\geq 20$ cMpc, and the results are shown in the panels of Fig.~\ref{Fig:profiles_z01234_COMbinsL}. Interestingly, for all bins, we observe the same trends with redshift as for the total population: the profiles are overall stable with time and show very little variation besides a slight increase of the densities at the cores.

These trends persist when restricting our analysis to the 5061 filaments with complete evolutionary tracks introduced in Sect.~\ref{Subsect:filament_merger_trees}. For the filaments at $z=0$ in a given bin of length, we compute the average profiles of their progenitors at higher redshifts, so that we track the radial evolution of the `same' filaments across cosmic time. The results of the progenitors are displayed by the coloured squares in Fig.~\ref{Fig:profiles_z01234_COMbinsL}. We observe quite a good agreement with the profiles of the general population and with the trends presented above.

For a more quantitative comparison we measure the difference $D$ between the average profiles ($p$) obtained at $z_1$ and at $z_2$, with respect to their corresponding errors ($e$), as follows:
\begin{equation}\label{Eq:difference}
    D(z_1, z_2) = \frac{p(z_1) - p(z_2)}{\sqrt{ e(z_1)^2 + e(z_2)^2} }.
\end{equation}
The differences between profiles of consecutive redshifts are reported in Fig.~\ref{Fig:profiles_variation_significance}, demonstrating that the variations in filament over-densities with time are indeed tiny ($< 2 \sigma$), even at filament cores.

\subsection{\label{SubSect:profiles_filament_populations}Comment on filament populations}

\begin{figure}
    \centering
    \includegraphics[width=0.5\textwidth]{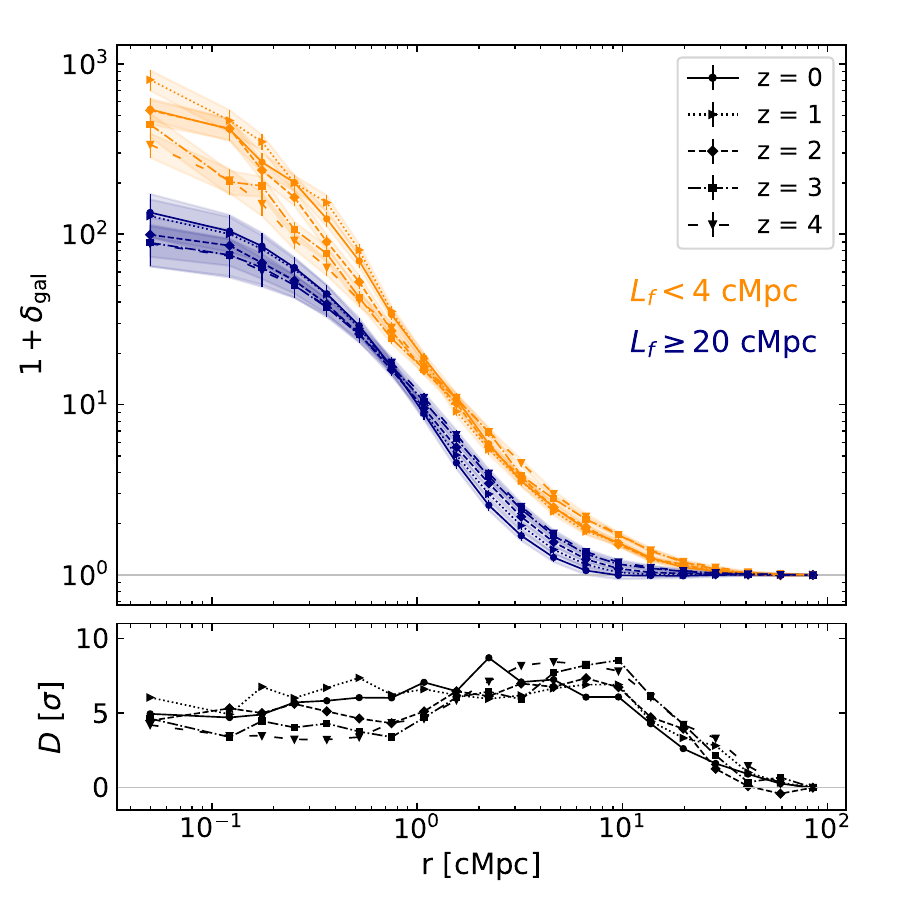}
    \caption{Evolution of density profiles of extreme bins of comoving length (top panel). The bottom panel presents the significance of the differences (adapted from Eq.~\ref{Eq:difference}) between the longest and shortest average filaments at a given redshift.}
    \label{Fig:profiles_z01234_SL}
\end{figure}

We finally compare the different panels of Fig.~\ref{Fig:profiles_z01234_COMbinsL} with each other. For all the redshifts it is remarkable to see that the shortest filaments have always higher density values than the longest ones. This trend is even more visible in Fig.~\ref{Fig:profiles_z01234_SL}, where we have over-plotted the profiles of the extreme populations at all $z$, in orange and blue for the shortest ($L_f < 4$ cMpc) and longest filaments ($L_f \geq 20$ cMpc), respectively. The bottom panel of this figure quantifies the difference $D$ (adapted from Eq.~\ref{Eq:difference}) between these two extremes at the same redshift. This evidences a persistent $\sim 5\sigma$ difference between the shortest and the longest populations of all redshifts. In addition, we can observe some differences in the overall shape of the profiles: those of the long population present flatter cores than those of the short one, which rather show steeper slopes at the innermost parts.

The results presented here complement the findings of \cite{GalarragaEspinosa2020, GalarragaEspinosa2021, GalarragaEspinosa2022} by demonstrating that the diversity of cosmic filaments observed at $z=0$ is also present at higher redshifts, with the same significance as at $z=0$. 
According to those studies, late time short filaments are preferentially located in the denser regions of the cosmic web \citep[e.g.~at the vinicity of nodes, see Fig.~1 of][]{GalarragaEspinosa2022} whereas long filaments inhabit less dense environments (e.g.~delimiting voids). 
These different locations in the cosmic web are most likely at the origin of the different comoving radial densities between short and long filaments, at all redshifts.

Finally, the density profiles presented in this work may include contributions from objects located at significant radial distances from a filament but which are most likely not physically associated with the filament. These might be, for example, galaxies belonging to an intersecting wall or void, or to other nodes. It might be interesting to separate the profiles of the filaments themselves from those characterising their environment, in a similar way as has been done by \cite{Ramsoy2021} for smaller-scale gaseous filaments. We expect to find significant differences between different populations of filaments, since the shortest structures are preferentially located in denser environments than the longest ones.

\section{\label{Sect:Conclusions}Summary, discussion, and conclusions}

We have used the hydro-dynamical runs of the MTNG simulation at redshifts $z=0$, 1, 2, 3, and 4 to build catalogues of cosmic filaments. We have analysed the spatial evolution of these structures across cosmic time, as well as the evolution of filament lengths, growth rates, and radial density profiles. 

Following an observational approach, we have used the most massive galaxies (in stellar mass) at each redshift as tracers of the density field (Sect.~\ref{SubSect:Choiceoftracers}), and extracted the cosmic skeleton using the DisPerSE filament finder. The full practical details related to running DisPerSE on discrete tracers were presented in Sect.~\ref{SubSubSect:Disperse101} with the aim of guiding future users through the steps of this code. 
By working at fixed number density of tracers we have implicitly imposed a scale for the detected filaments. This allowed us to perform our study on the large-scale cosmic structures without the contamination of smaller-scale filaments (such as, e.g. filaments linking satellites with their central galaxies). 

An important outcome of this work is the introduction of a method based on physical priors that allows us to test and quantify the robustness of the cosmic web skeletons that we identify. This was presented in full detail in Sect.~\ref{SubSubSect:Dispersecalibration} with the hope that future work in this field will address these questions with similar care. Keeping these considerations in mind can help the community move towards studies that are less dependent on the specific techniques used to detect filaments in the cosmic web.\\

We have quantified the growth of proper filament lengths and radial extensions in Figs.~\ref{Fig:filament_length_function} (left) and \ref{Fig:APP_profiles_all_z01234_proper}. This growth in proper coordinates is expected for structures at these large scales as a result of the cosmic expansion. 
The most interesting conclusions have however arisen when working in comoving coordinates, once the Hubble flow is factored out and similar types of filament can be compared at different redshifts. 
These conclusions are summarised and discussed in the following:

(i) We have found surprisingly stable filament length functions (Fig.~\ref{Fig:filament_length_function}, right) and over-density profiles (Figs.~\ref{Fig:profiles_all_z01234}) across redshift, demonstrating very little evolution of the global population of cosmic filaments over the past $\sim 12.25$ Gyrs. These results complement those of \cite{GhellerVazza2016}, who have shown only mild variations of filament scaling relations (e.g.~mass vs temperature) from redshift 0 to 1. The very small redshift dependence of filament over-densities found in this work shows that filaments are not becoming thicker or thinner with time in comoving coordinates. This could be directly related to the continuous accretion of matter from walls or voids, thus building up mass with time, and compensating for the flow of matter along the filaments towards nodes. Indeed, filaments can act as highways for matter transport as suggested by the theory of large-scale structure formation \citep{Zeldovich1970, Peebles1980, Zeldovich1982Nature, Bond1996}. Yet, to our knowledge, direct and quantitative evidence for this picture is still missing.

(ii) Going beyond the study of global trends, we have associated filaments across redshifts and built their evolutionary tracks from $z=0$ to 4 in Sect.~\ref{Subsect:filament_merger_trees}. This filament association, performed for the first time in this work, allowed us to follow the evolution of a relatively small but significant number of structures across cosmic time. By tracking individual filaments we have shown that some of them actually significantly change their length with time (they can contract or elongate, Fig.~\ref{Fig:growth_rate_vs_L}), while the properties of the global population are preserved (see the point above).
We have quantified the expansion or contraction along the filament's longitudinal axis by computing growth factors $\Gamma$ and $\Gamma_n$ (Figs.~\ref{Fig:growth_rate_distrib} and \ref{Fig:growth_rate_vs_L}) finding that, at all times, filament expansion along the longitudinal direction is related with length. Indeed, we have shown that filaments longer than a certain length are preferentially expanding whereas the shortest structures always collapse. We anticipate that further characterising these trends by modelling the filament growth in, for example, different DE scenarios might open doors towards new cosmological probes.

(iii) Conclusions regarding the diversity of cosmic filaments at a given time have also been reached. At a fixed redshift, filament length distributions span more than two orders of magnitude in comoving megaparsecs (Fig.~\ref{Fig:filament_length_function}, right panel), and the over-density profiles of populations of extreme lengths, i.e.~the short and long filaments, always differ by $\sim 5\sigma$ (Fig.~\ref{Fig:profiles_z01234_SL}). In the light of previous work \citep{GalarragaEspinosa2020, GalarragaEspinosa2021, GalarragaEspinosa2022}, we argue that the existence of this diversity of filaments at all redshifts is likely to arise from environmental effects. Different populations inhabit different locations within the cosmic web, where either gravity or DE dominate the local evolution. Within this picture, high redshift cosmic filaments located in high density regions would be more likely to collapse and become part of the short population at later times. Conversely, if the same high-$z$ filaments are found in less dense environments (e.g.~next to cosmic voids), they would rather expand, due to the stretching of the cosmic web caused by the DE forces.

Future perspectives for cosmic filament studies beyond those already mentioned throughout the paper lie in the analysis of the evolution of gas properties. In particular, investigating the emergence of the WHIM phase, which dominates the gas fraction ($> 80 \%$) of cosmic filaments at $z=0$, can be insightful for the study of the late-time missing baryons. The evolution of the cold gas phase and its relation with filamentary structures should also be further explored to improve our understanding of galaxy evolution.

In conclusion, the results presented in this work are an important step forward in the characterisation of cosmic filaments. With further theoretical work these structures can potentially become tools for cosmological studies, supplementing their well-established importance in the galaxy evolution field. Better understanding cosmic filaments is crucial to fully exploit the potential of this decade's experiments (such as the already mentioned DESI, Euclid, JPAS, Vera Rubin, or PFS galaxy surveys) but also the potential of on-going and future experiments focusing on the detection of gas in different phases, such as CLAMATO, eROSITA, Athena, and the Simons Observatory.

\section*{Data availability}

The MillenniumTNG simulations will be publicly released at \url{https://www.mtng-project.org} in the near future. The cosmic filament catalogues constructed and analysed in this work will be made available to the community at the same webpage.

\section*{Author contributions}
DGE led the project: project conceptualisation, methodology, software, validation, formal analysis, investigation, data curation, writing (original draft, review \& editing), and visualisation.
CC helped with project conceptualisation, methodology, software, writing (review \& editing). 
CG helped with project conceptualisation and writing (review \& editing). 
SW helped with methodology, writing (review \& editing), and resources.
VS and RP helped with data curation, writing (review \& editing), and resources.
BH, SB, FF, RK, LH helped with writing (review \& editing), and resources. MB, AMD, CHA helped with resource provision.

\begin{acknowledgements}
We thank the anonymous referee for their interesting and insightful comments. DGE would like to thank Charlotte Welker, Marcelo Musso, Guinevere Kauffmann, Toni Tuominen, Enrico Garaldi, Fabrizio Arrigoni-Battaia, Joanne Cohn, and the ByoPiC team\footnote{\url{https://byopic.eu/}} for insightful comments and discussions during the preparation of this work.
This research was supported in part by the National Science Foundation under Grant No. NSF PHY-1748958, and a fraction of it was carried out during the KITP Cosmic Web workshop. SB is supported by the UK Research and Innovation (UKRI) Future Leaders Fellowship [grant number MR/V023381/1]. CC acknowledges support from the Knut and Alice Wallenberg Foundation and the Swedish Research Council (grant 2019-04659). CG acknowledges support from the European Research Council (ERC) under the European Union’s Horizon 2020 research and innovation programme grant agreement ERC-2015-AdG 695561 (ByoPiC project). 

\end{acknowledgements}

\bibliography{main} 


\begin{appendix}

\section{\label{Appendix:first_progenitors}Spatial distribution of first progenitors of the subhaloes at $z=0$}

\begin{figure*}[h!]
    \centering
    \includegraphics[width=1\textwidth]{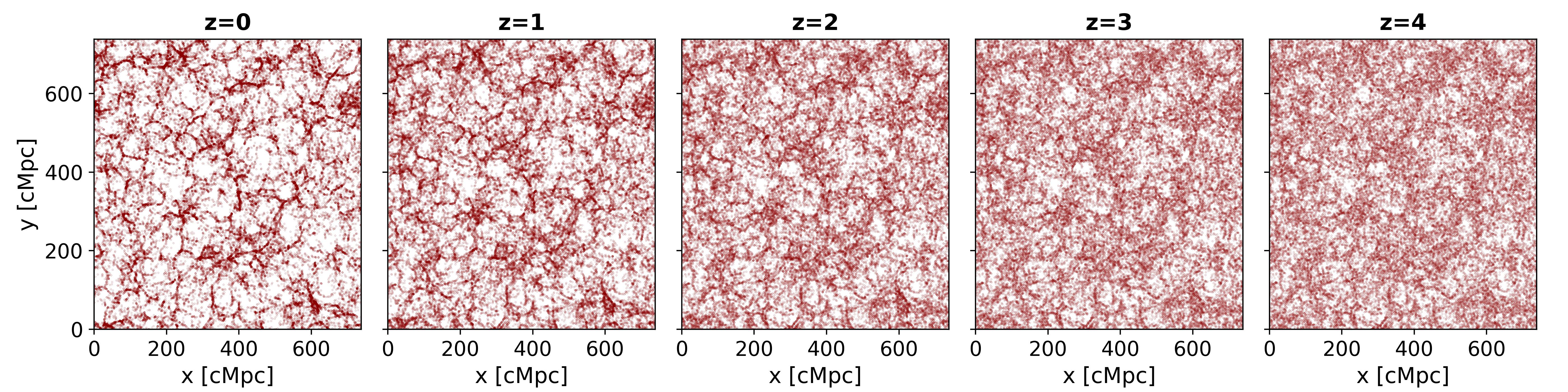}
    \caption{Spatial distribution at different redshifts of the first progenitors of the subhaloes of stellar masses $\log_{10}(M_* / \mathrm{M}_\odot) \geq 9$ at $z=0$. The different panels show the same slice of thickness 15 cMpc as the one in Fig.~\ref{Fig:slice_gals_fils}.}
    \label{Fig:APP_first_progenitors}
\end{figure*}

Figure~\ref{Fig:APP_first_progenitors} shows, for the different redshifts explored in this work, the spatial distribution of the first progenitors of the $z=0$ galaxies of stellar masses $\log_{10}(M_* / \mathrm{M}_\odot) \geq 9$. It is clearly apparent that these galaxies are not optimal tracers of the cosmic skeleton. Indeed, as we move to higher redshift, the filamentary pattern (finely delineated at $z=0$) becomes noisier, hard to distinguish at $z=3$, and absent by $z=4$. This figure can be compared with Fig.~\ref{Fig:slice_gals_fils} of the main text, which shows, in the same slice, that the more massive galaxies at any redshift are better tracers of the cosmic skeleton.
More quantitatively, we find that the first progenitors of the selected $z=0$ galaxies have stellar masses as low as $\log_{10}(M_* / \mathrm{M}_\odot) = 6.47$ at $z=1$. By $z=4$, only $76 \%$ of them have a non zero stellar component, the rest being only low mass DM clumps.

\section{\label{Appendix:matchingCPmax_FOF}Matching the CPmax with massive haloes at $z=0$}

The full results of the matching of CPmax with FoF haloes are presented in Table~\ref{Table:APP_match_CPmax_Clusters}, together with all the details regarding the number of filaments, CPmax, and length statistics of each of the eight skeletons extracted at $z=0$ with different DisPerSE parameterisation.

\begin{table*}[h!]
\caption{Details of DisPerSE skeletons extracted with different parameterisation at $z=0$. Note: results after cleaning the skeletons.}
\label{Table:APP_match_CPmax_Clusters}   
\centering  
\begin{tabular}{ l | c  c  c  c | c  c  c  c }
\hline \hline
    \rhodtfe~smoothing & S & S & S & S & noS & noS & noS & noS \\
    Persistence & $1\sigma$ & $2\sigma$ & $3\sigma$ & $4\sigma$ & $1\sigma$ & $2\sigma$ & $3\sigma$ & $4\sigma$ \\ 
     \hline
       Number filaments & 318 337 & 169 497 & 114 269 & 75 065 &
  697 234 & 395 033 & 255 200 & 172 280 \\
  Number CPmax & 61 035 & 41 064 & 29 600 & 18 157 &
  173 370 & 114 399 & 76 777 & 51 040 \\
  Min length [Mpc] & 0.0001 & 0.20 & 0.23 & 0.23 &
  0.0001 & 0.001 & 0.003 & 0.03 \\
  Max length [Mpc] & 82.16 & 115.61 & 127.41 & 146.77 &
  72.0 & 71.5 & 105.5 & 147.4 \\
  Mean length [Mpc] & 10.5 & 12.3 & 14.7 & 20.6 &
  6.4 & 7.3 & 8.8 & 11.0 \\
  Median length [Mpc] & 8.9 & 9.8 & 11.3 & 15.9 &
  5.1 & 5.5 & 6.3 & 7.9 \\
    \hline 
  Galactic haloes hosting CPmax & $2.0\%$ & $0.9\%$ & $0.51\%$ & $0.25\%$  
  & $10.1\%$ & $5.9\%$ & $3.1\%$ & $1.5\%$ \\
  Groups hosting CPmax & $48.3\%$ & $36.4\%$ & $27.3\%$ & $16.9\%$ 
   & $74.2\%$ & $65.3\%$ & $55.4\%$ & $43.4\%$\\
  Clusters hosting CPmax & $97.0\%$ & $91.3\%$ & $80.9\%$ & $59.1\%$
   & $99.7\%$ & $98.7\%$ & $96.7\%$ & $91.3\%$ \\
  \hline
  CPmax in galactic haloes & $21.6\%$ & $14.6\%$ & $11.2\%$ & $8.7\%$  
  & $37.5\%$ & $33.2\%$ & $26.3\%$ & $18.7\%$ \\
  CPmax in Groups & $55.7\%$ & $62.3\%$ & $64.8\%$ & $65.6\%$ 
   & $30.4\%$ & $40.3\%$ & $50.9\%$ & $59.9\%$ \\
  CPmax in Clusters & $9.9\%$ & $13.7\%$ & $16.7\%$ & $20.0\%$ 
   & $4.7\%$ & $5.7\%$ & $7.8\%$ & $11.0\%$ \\
  Total fraction of matched CPmax & $87.1\%$ & $90.6\%$ & $92.8\%$ & $94.4\%$ 
   & $72.7\%$ & $79.1\%$ & $84.9\%$ & $89.6\%$
   \\ 
    \hline
  Comments & impure & ok & incomplete & incomplete
  & impure & impure & impure & pb lengths \\
  \hline
 \end{tabular}
\end{table*}

\section{\label{Appendix:Profiles}Density profiles in proper radial coordinates }

\begin{figure*}[h!]
    \centering
    \includegraphics[width=0.35\textwidth]{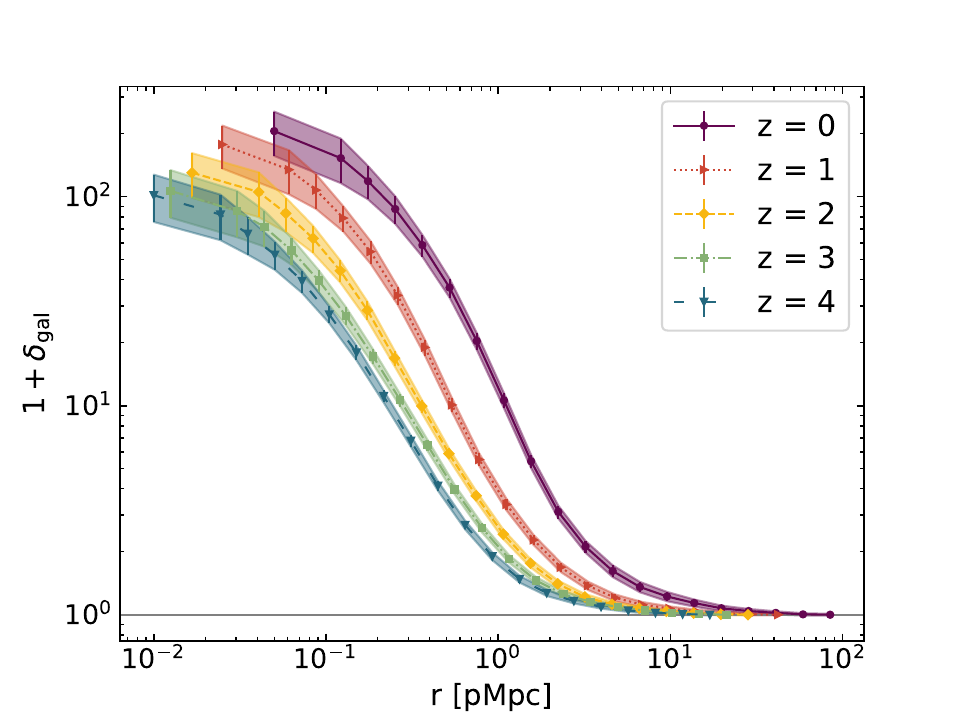}
    \caption{Same as Fig.~\ref{Fig:profiles_all_z01234} but with $r$ in units of proper megaparsec.}
    \label{Fig:APP_profiles_all_z01234_proper}
\end{figure*}

\begin{figure*}[h!]
    \centering
    \includegraphics[width=0.75\textwidth]{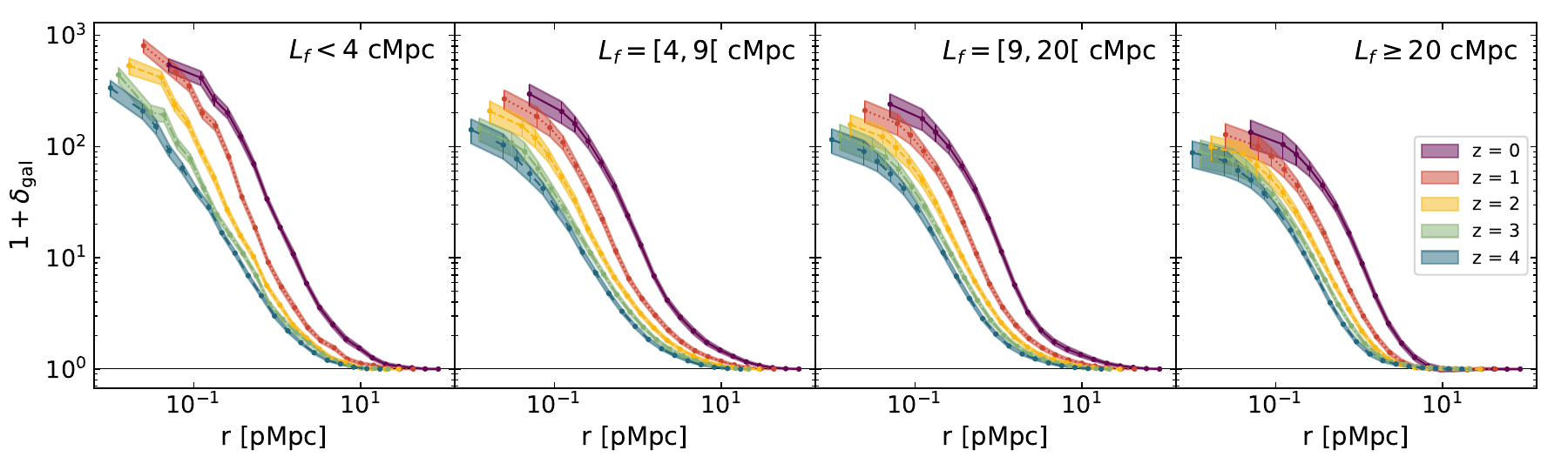}
    \caption{Same as Fig.~\ref{Fig:profiles_z01234_COMbinsL} but with $r$ in units of proper megaparsec.}
    \label{Fig:APP_profiles_binsL_z01234_proper}
\end{figure*}

In proper coordinates, we observe the expected trends that the density values and radial extensions of filaments increase with time. This is shown by Figs.~\ref{Fig:APP_profiles_all_z01234_proper} and \ref{Fig:APP_profiles_binsL_z01234_proper} of this appendix. 

\FloatBarrier

\end{appendix}

\end{document}